\documentclass[12pt,a4paper]{article}
\usepackage{times}
\usepackage{graphicx}
\usepackage{placeins}
\usepackage[export]{adjustbox}
\usepackage{ifxetex}
\usepackage{svg}
\usepackage{amsthm}
\usepackage{caption}
\usepackage[margin=1in,footskip=0.5in]{geometry}
\usepackage{setspace}
\usepackage[]{amsmath}
\usepackage{amsfonts}
\usepackage{sectsty}
\usepackage[utf8]{inputenc}
\usepackage{xcolor}
\usepackage[space]{grffile} 
\usepackage{epstopdf}
\usepackage{subfloat}
\usepackage{subfig}
\usepackage[space]{grffile}
\usepackage{xspace}
\usepackage{mathtools}
\usepackage{bbold}
\usepackage[labelfont=bf,font=small]{caption}
\usepackage{multirow,array}
\usepackage{parskip}
\newtheorem{definition}{Definition}
\newtheorem{proposition}{Proposition}
\newtheorem{theorem}{Theorem}
\newtheorem{lemma}{Lemma}
\newtheorem{corollary}{Corollary}

\newtheorem*{proposition01}{Proposition (Singh et al.)}

\newtheorem{assumption}{Assumption}

\newcommand{\inlinetitle}[1]{\noindent\textbf{#1}\\}

\newcommand\blfootnote[1]{%
  \begingroup
  \renewcommand\thefootnote{}\footnote{#1}%
  \addtocounter{footnote}{-1}%
  \endgroup
}
\title{Algorithmic collusion with endogenous exploration}
\date{\today}
\author{Ivan Conjeaud$\dagger$}
\begin{document}

\maketitle
\begin{abstract}
    I study a two-stage model in which two players simultaneously choose an exploration parameter for their Q-learning algorithms, which then repeatedly play a one-shot game chosen from a class of social dilemmas including the prisoner's dilemma, first and second-price auctions as well as Bertrand competition with horizontally differentiated products. The players collect the limit average payoffs obtained by their algorithms. I show that all equilibria are collusive: both players receive payoffs that are strictly higher than the payoffs received in the unique strict Nash equilibrium of the one-shot game. I then use extensive numerical simulations in a Bertrand duopoly and a parameterized prisoner's dilemma. Their results allow to gain insight on (i) the mechanism causing algorithmic collusion and (ii) the strategic role of exploration levels in the  game. They reveal that in equilibrium, the players tend to choose algorithms that \textit{over-explore}, which comes at the detriment of joint payoff. These findings have important implications for algorithmic collusion.
    \medbreak
\end{abstract}
\blfootnote{$\dagger$ Affiliations: \textit{Paris
School of Economics, Université Paris 1 Panthéon Sorbonne}, 48 Boulevard Jourdan, 75014 Paris. \textit{Aix Marseille Université, CNRS, AMSE} Marseille, France. \textit{Contact:} ivan.conjeaud@psemail.eu}
\newpage
\section{Introduction}
Algorithms are increasingly used to take decisions on behalf of humans in a wide range of contexts. Reinforcement learning algorithms have raised concerns as they may learn to collude tacitly. Previous theoretical work has confirmed that reinforcement learning algorithms—most prominently Q-learning—can sustain collusive outcomes in repeated interactions. Yet, their ability to sustain collusive outcomes hinges on parameters that govern how the algorithms explore their environment, parameters that the literature has so far treated as exogenous. This paper takes a step further by endogenizing the choice of exploration. Before playing, each agent strategically selects the exploration rate of their Q-learning algorithm. I show that when exploration rates are chosen optimally by the agents operating the algorithms, collusion inevitably emerges in equilibrium. At the same time, exploration rates can be used strategically, which tends to soften collusion but never fully eliminates it. 
\medbreak
A rapidly expanding literature has been investigating whether reinforcement learning can generate collusive behavior in repeated games. Calvano et al. (2020) \cite{Calvano2020} show numerically that independent Q-learning algorithms can learn to sustain supra-competitive prices by learning punishment-reward schemes, even without explicit communication. Subsequent work has explored the robustness of this phenomenon to different types of reinforcement learning algorithms and economic environments (e.g., Klein, 2021 \cite{Klein2021}; Asker et al. 2022 \cite{Asker2022}; Calvano et al. 2021 \cite{Calvano2021}; Hettich, 2021 \cite{hettich2021algorithmic}). A common feature of this literature is that the parameters governing exploration—the extent to which algorithms experiment with new actions—are imposed exogenously. These parameters, however, play a central role in determining whether algorithms engage in tacit collusion. It has been shown by Banchio and Mantegazza (2022) \cite{banchio_mantegazza} that Q-learning algorithms with an $\varepsilon$-greedy policy playing a parameterized prisoner's dilemma engage in mutual cooperation when their (symmetric) exploration rate is low enough. When the exploration rate is too high, the cooperative behavior disappears, and the algorithms learn to defect. By treating exploration as a strategic choice rather than a fixed feature of the algorithm, this paper shows that collusion is an equilibrium outcome. It highlights that the concerns about algorithmic collusion cannot be alleviated by merely letting the firms operating the algorithms optimally select their parameters.
\medbreak
This paper introduces a two-stage game in which two agents endogenously select the exploration rates of their Q-learning algorithms with $\varepsilon$-greedy policy prior to letting them engage in a repeated interaction. I show that when agents anticipate the long-run performance of their algorithms, they optimally select exploration parameters that generate algorithmic collusion. In order to do so, I characterize the long-run outcome reached by the algorithms in two special cases. First, when one algorithm is \textit{greedy} (i.e. never explores) while the other one chooses a positive exploration rate, no collusion is generated. The greedy algorithm eventually settles for the competitive action (defection in the prisoner's dilemma, bidding its valuation in an auction and setting the competitive price in Bertrand pricing games). The other plays the competitive action \textit{most of the time} but since it uses a positive exploration rate, it also sometimes plays cooperatively. By doing so, it receives a payoff that is lower than the one received in the competitive equilibrium of the one-shot game, while the greedy algorithm collects a payoff strictly higher. Second, when both algorithms are greedy, some collusion can be generated and for both algorithms the limit of the average payoffs they collect needs to be higher than the payoff in the competitive equilibrium of the one-shot game. The characterization of the behavior of the greedy algorithm reveals an important feature of Q-learning, that allows it to secure strictly supra-competitive payoffs. The Q-learning algorithm enforces a form of \textit{reactive commitment to cooperation}. It leaves scope for mutual cooperation, but reacts by settling to uncooperative actions when the other player's algorithm tries to abuse it. As a consequence, the agents are always able to sustain payoff that are strictly higher than the ones received in the competitive equilibrium of the one-shot game. I then use extensive numerical simulations in the case of a Bertrand duopoly and of a parameterized prisoner's dilemma to uncover the strategic role of exploration rates. I use these to (i) detect and measure the intensity of collusion for different values of the parameters and (ii) computationally find the Nash equilibria of the two-stage game. For appropriate parameters, the Q-learning algorithms alternate between phases of mutual cooperation and mutual defection, with asymmetric phases in between, during which one algorithm is more cooperative than the other. The numerical simulations reveal that increasing one's exploration level allows one to (i) play defection more often in cooperative phases and (ii) lengthen and increase the frequency of \textit{favorably asymmetric phases}, which comes at the expense of overall cooperation. This creates \textit{over-exploration} in equilibrium. Hence the endogenous choice of exploration levels softens collusion, but does not completely erase it.
\medbreak
The rest of the paper is organized as follows. Section \ref{sec: lit review} reviews the literature related to the present work, Section \ref{sec: model} formally introduces the model, Section \ref{sec: collusion in eq} states the main result and describes the proof, Section \ref{sec: simulations} describes the numerical experiments and present their results, Section \ref{sec: conclusion} concludes.
\section{Related literature} \label{sec: lit review}
The present article relates to a broad literature on Multi Agent Reinforcement Learning (MARL), which lies at the interface of computer science and game theory, and is mainly concerned with the interaction of several reinforcement learning algorithms. An introduction to this literature is given by Nowé et al. (2012) \cite{nowe2012game}. MARL presents the difficulty of having very few theoretical guarantees on the convergence of algorithms. Important contributions in the field have focused on designing algorithms that extend classical Q-learning, such as Tesauro (2003) \cite{tesauro2003extending} with Hyper Q-learning and Hu (2003) \cite{hu2003nash} with Nash Q-learning. Other authors have focused on building algorithms that maintain cooperation in environments resembling the prisoner's dilemma by adapting modern reinforcement learning techniques, such as Lerer and Peysakhovich (2017) \cite{lerer2017maintaining} and Tampuu et al. (2017) \cite{Tampuu2017}. Closer to the current work, Kianercy and Galstyan (2012) \cite{kianercy2012dynamics} provide a full characterization of the rest points of a system composed of two Q-learning algorithms with Boltzmann exploration policy playing $2\times 2$ games. Generally, it is well-known that some algorithms tend to learn cooperation when repeatedly playing a prisoner's dilemma, see Banerjee and Sen (2007) \cite{banerjee2007reaching} for an example. A comprehensive survey of the MARL literature has been provided by Busoniu et al. (2008) \cite{busoniu2008comprehensive}.\medbreak This paper also relates to a vast literature on \textit{supergames} of the prisoner's dilemma, as termed by Friedman (1971) \cite{friedman1971non}. In such supergames, players choose a strategy of the repeated prisoner's dilemma, that prescribes what action to select as a function of the history of play. Fundamental contributions in this field have been produced by Axelrod (1980, 1981) \cite{axelrod1980effective,axelrod1981emergence} who highlighted the properties of tit-for-tat. Tit-for-tat (TFT) prescribes to play cooperatively initially and to subsequently mimic the opponent's last action. This strategy has proven very effective in sustaining cooperation by enforcing a form of \textit{reactive commitment}. An opponent to TFT is left space to cooperate, yet is immediately punished in case of defection. The results of the present paper highlight a similar property to Q-learning algorithms, which allow to sustain some cooperation by reactively committing to cooperation. To the difference of TFT however, Q-learning requires no initial bias towards cooperation. An enormous body of research has investigated the so called "Folk Theorem", stating that feasible and individually rational payoffs of a one-shot game can be realized as an average payoff in some equilibrium of the supergame. The seminal contribution in this literature is that of Fudenberg and Maskin (1986) \cite{fudenberg1986folk}, and has been followed by countless other articles.  Recent contributions have bridged folk theorems with reinforcement learning algorithms, as those of Askenazi-Golan et al. (2024) \cite{askenazi2024reinforcement} and Cartea et al. (2022) \cite{cartea2022algorithms}. An important literature has also investigated supergames of the prisoner's dilemma restricted to simple algorithms. In Rubinstein (1986) \cite{rubinstein1986finite}, agents choose a finite state automaton to repeatedly play a prisoner's dilemma on their behalf. The author proves that, if agents incur a cost for maintaining the complexity of their machines, cooperation is never an equilibrium outcome (where equilibrium refers to \textit{semi-perfect equilibrium}, as defined in the article). This consideration on the cost of maintaining the machine complexity is neglected in the present contribution, given the simplicity of Q-learning algorithms and the low computational costs of modern machines. Other similar contributions include that of Miller (1996) \cite{miller1996coevolution}, who takes an evolutionary perspective and that of Cho (1995) \cite{cho1995perceptrons}, who studies a repeated prisoner's dilemma played by perceptrons (the simplest form of neural networks). 

Two features differentiate the main result of this article from a folk theorem. First, the equilibria of the supergame yield payoffs that are \textit{strictly} individually rational rather than weakly. This rules out that the Nash equilibrium of the one-shot game considered be reproduced by the means of algorithms with strategically chosen exploration levels. Second, it does not provide an existence guarantee for individually rational payoffs. In the folk theorem literature, such a result is generally obtained by directly constructing equilibrium algorithms that generate the desired vector of payoffs. A construction of this type is not possible with Q-learning algorithms, as their behavior for general couples of exploration parameters is impossible to describe analytically. \medbreak This work contributes more specifically to the literature on algorithmic collusion, which studies the interactions between Q-learning algorithms in economic situation. The possibility of spontaneous collusion by Q-learning algorithms has been receiving important attention in recent years. On the empirical side, Assad et al., 2020 \cite{Assad2020} have provided evidence for algorithmic collusion on the German gasoline retail market, after algorithmic pricing methods became widely available in 2017. A similar phenomenon has been highlighted by Musolff et al., 2022 \cite{Musolff2022} using Amazon data. The policy implications and practical relevance of the topic are discussed in Calvano et al., 2019 \cite{Calvano2019}. On the theoretical side, the seminal contributions are those of Calvano et al., 2020 \cite{Calvano2020} and of Klein, 2021 \cite{Klein2021}. Using extensive simulations, they show that Q-learning algorithms playing a repeated pricing game (in a Bertrand oligopoly in the first case, in a setting \textit{à la} Maskin and Tirole, 1988 \cite{maskin1988theory} in the second) learn to set supra-competitive prices in the long run. Calvano et al. (2020) emphasize the \textit{anatomy of collusion} constituted by punishment-reward schemes upon deviation, which is made possible by letting algorithms condition their play on the previous action played by their opponent. In Klein (2021) on the other hand, long run behavior is characterized by asymmetric cycling prices rotating demand between players. Simulation-based extensions have followed. Calvano et al., 2021 \cite{Calvano2021} consider a setting with a Cournot duopoly with stochastic demand and show that a similar collusive behavior appears. Colliard et al. (2022) \cite{colliard2022algorithmic} investigate the behavior of interacting Q-learning algorithms (\textit{algorithmic market makers}) setting prices for a risky asset. Banchio and Skrzypacz, 2022 \cite{banchio2022artificial} consider Q-learning algorithms repeatedly playing classical auction games, and interestingly report a difference between second-price and first-price auctions. In second-price auctions, algorithms converge to the static equilibrium of the one-shot game, while in first-price auction collusive behavior appears.  Hettich, 2021 \cite{hettich2021algorithmic} runs simulations using a more advanced technology (namely deep Q-learning, see Mnih et al., 2015 \cite{mnih2015human} for a description of this deep reinforcement learning method) and obtains results similar to those of Calvano et al. (2020) with a faster convergence to collusive behavior. Further extensions have been considered, typically by investigating the role of different market structures as in Sanchez-Cartas and Katsamakas, 2022 \cite{SanchezCartas2022}, or in Abada and Lambin, 2022 \cite{abada2023artificial}, which find similar results to Calvano et al. (2020) on an economic environment replicating electricity markets. Further, Johnson et al., 2023 \cite{johnson2023platform} study the effect of platform design on the behavior of Q-learning algorithms. They show that some platform designs that are effective for classical players turn out to be socially harmful in the presence of Q-learning algorithms, and point out a better adapted one.
\medbreak
Another branch of research, which the present work mostly builds on, has focused on understanding the mechanism responsible for collusion, which relies on the asynchronous nature of Q-learning's updating. By simulating stateless Q-learning algorithms with $\varepsilon$-greedy policies, Asker et al., 2022 \cite{Asker2022} provide evidence that collusive behavior is rooted in Q-learning's asynchronous updating. They highlight that if algorithms have access to minimal information and are given minimal economic reasoning (specifically the demand being downward sloping), then collusion is substantially reduced. A similar phenomenon is observed by Banchio and Skrzypacz (2022): letting Q-learning have access to the highest bid in previous period and letting them compute counterfactual scenarios when updating Q-values precludes collusive behavior. Finally, Banchio and Mantegazza (2023) \cite{banchio_mantegazza}, characterize the limiting behavior of Q-learning algorithms playing a repeated prisoner's dilemma using continuous time approximations. More precisely, they provide a theoretical bound on exploration level under which collusive behavior is possible\footnote{A similar result was pointed out by Abada and Lambin (2022) using simulations}. They highlight a phenomenon of \textit{spontaneous coupling} between $\varepsilon$-greedy algorithms, and formally prove that letting algorithms synchronously update the Q-values prevents collusion. The second part of this paper extends their results by means of numerical simulations.
\medbreak
The present article aims to fill a gap in the existing literature on algorithmic collusion, in which studies of competition on the algorithms are rare. Brown and MacKay (2023) \cite{brown2023competition} study a setting in which two firms strategically choose a pricing algorithm, and more specifically highlight the impact of asymmetric update frequency on the equilibrium prices. They show how the use of simple "algorithms", which in their article are simply functions mapping the latest current payoff-relevant price posted by the opponent's algorithm to prices, lead to supra-competitive outcomes when the two players have asymmetric pricing technologies. They also provide an analysis of equilibrium collusive pricing when both firms can price "in real time". In such a case, by conditioning on the other's price, a firm can induce its opponent to increase its price, leading to collusive outcome. In this article, I take a different approach. The algorithms studied in the present article are much more complex (even though they remain the simplest form of reinforcement learning) and, more importantly, they do not allow to directly condition on the opponent's past or current price. Reinforcement learning is thus much less demanding in terms of the understanding players need to have of their environment. With extensive simulations in the case of the prisoner's dilemma, I will show in Sec. \ref{sec: simulations} how asymmetry in the exploration policies is detrimental to collusion. Adopting a higher exploration policy than one's opponent allows one to play defection more often while the other's algorithm keeps cooperating. This is due to the fact that a high exploration rate enables an algorithm to realize faster that defection is a dominant action, which allows one to exploit the other's slower algorithm before he realizes as well. This provides an incentive to unilaterally raise their exploration rates, which hampers collusion in equilibrium. \medbreak In the context of algorithmic collusion, having different algorithms face each other has been investigated by Sanchez-Cartas and Katsamakas (2022), who compare Q-learning with Particle Swarm Optimization (PSO). PSO (Kennedy and Eberhart 1995 \cite{kennedy1995particle}) is a meta-heuristic method of optimization belonging to the class of evolutionary algorithms. Given a function to minimize (without knowing its gradient), PSO simulates the evolution of \textit{candidate solutions} (seen as particles in a swarm). The particles move in the solution space depending on their own best known solution and the best known solution of the swarm, and oftentimes manage to collectively find optima. Their simulations indicate that, when PSO is competing with a stateless Q-learning algorithm, both set supra-competitive prices, however equilibrium considerations are absent and the choice of the parameters in the Q-learning algorithm and the choice of PSO is exogenous. So far, only Compte (2023) \cite{compte} integrates equilibrium considerations. In this paper, a variation of Q-learning integrating a possible bias towards cooperation is considered. Biases are chosen simultaneously by players before the algorithms start running. Simulations using a prisoner's dilemma indicate that Nash equilibria featuring positive bias towards cooperation exist and enhance collusive behavior. Finally, recent work by Dolgopolov (2024) \cite{dolgopolov2024reinforcement} and Xu and Zhao (2024) \cite{xu2024mechanism} have provided further insight into the behavior of interacting Q-learning algorithms. Dolgopolov (2024) provides a full characterization of stochastically stable states of Q-learning algorithms repeatedly playing a prisoner's dilemma. He proves that under $\varepsilon$-greedy policies, the only stochastically stable outcome is for both algorithms to play defection, while depending on the payoffs and the learning rate, logit exploration allows for some cooperative behavior. In a similar way, Xu and Zhao (2024) prove that, for a class of games included in the one studied in the present paper, in stochastically stable outcomes algorithms learn to play the strict Nash equilibrium of the game. They then give insight on why there is a difference between what stochastic stability allows and the empirical behavior of Q-learning algorithms using numerical simulations. While they allow for asymmetric exploration rates, these two contributions assume that the exploration rates vanish and focus on the limiting behavior of the algorithms. By contrast, the present article assumes that exploration rates are constant and focuses on the strategic use of exploration levels rather than on the limiting behavior of the algorithms per se. 
\section{The model} \label{sec: model}
Two players, $A$ and $B$, use memory-less Q-learning algorithms to repeatedly play a one-shot game on their behalf. In a first stage $A$ and $B$ simultaneously select an exploration parameter to be implemented by a Q-learning algorithm with $\varepsilon$-greedy policy. In a second stage, their algorithms repeatedly play the one-shot game. The players evaluate the payoffs derived from their exploration policy according to limit-of-means criterion.
\subsection{One-shot game} \label{sec: One-shot game}
Denote $G_0=\Big((A,B) \ , (A_A,A_B) \ , (u_A, u_B) \Big)$ the one-shot game where $A_i$ denotes the action set of player $i$ and $u_i: A_i \times A_{-i} \longrightarrow \mathbb{R}$ denotes his payoff function. We take the following assumptions on $G_0$.
\begin{enumerate}
    \item $G_0$ is symmetric and has $K \in \mathbb{N}^*$ actions. 
    \item The action set is indexed such that $A_A=A_B=\mathcal{A}=\{a_1,...,a_K\}$ and for $n>k$:
    \begin{enumerate}
        \item $u(a_n,a_n)>u(a_{k}, a_k)$
        \item for all $1\le m \le K$, $u(a_m, a_n)\ge u(a_m, a_k)$
    \end{enumerate}
    \item $(a_1, a_1)$ is a strict Nash equilibrium.
\end{enumerate}
The set of games satisfying these assumptions is denoted $\mathcal{G}$. The crucial feature of these games is that their actions are indexed by a degree of cooperativeness. In symmetric profiles, increasing the level of cooperation increases the joint payoff $(2.a)$. Regardless of their action, players are better off when their opponent cooperates more $(2.b)$. However the symmetric profile with the smallest level of cooperation is a strict Nash equilibrium, so that $G_0$ is a social dilemma. Among the games that satisfy these conditions, three examples are particularly relevant.
\medbreak
\textbf{Example 1 (prisoner's dilemma) } In the prisoner's dilemma, there are two actions. Let $a_1$ be defection and $a_2$ be cooperation. The payoffs are ordered as follows: $u(a_1, a_2) > u(a_2,a_2) > u(a_1, a_1)>u(a_2, a_1)$. The payoff obtained under mutual cooperation is higher than the payoff obtained under mutual defection (assumption $2. \ a$). A player is always better off if his opponent cooperates rather than defects (assumption $2. \ b$) and mutual defection is the only (strict) Nash equilibrium. 
\medbreak
\textbf{Example 2 (discretized first-price auction)} Consider a first-price auction with two bidders, $A$ and $B$, who have the same valuation $v$. Assume they select their bids from the same discrete grid $\{v-bk, k\in \{1,...,K\}\}$ where $b<v$ and $v-bK\ge 0$. Under the profile $(b_i,b_j)$ the expected payoff to firm $i$ is: \begin{equation}
    u(b_i,b_j)=\begin{cases}
      v-b_i & \text{if } b_i>b_j\\
      \frac{v-b_i}{2} & \text{if } b_i=b_j\\
      0 & \text{if } b_i<b_j
    \end{cases}
\end{equation}
In this game, $b_i$ is more cooperative than $b_j$ if $b_i<b_j$. Under symmetric profiles, the payoffs to players increase as their bids decrease (assumption $2 \ a$). If a player places bid $b_i$, then as the bid of the opponent decreases his payoff goes from $0$ to $\frac{v-b_i}{2}$ and eventually to $v-b_i$ (assumption $2 \ b$). Finally, under the profile $(b_K, b_K)$ a unilateral deviation would lead to losing the auction and making payoff $0$, so that this profile is a strict Nash equilibrium (assumption $3$).
\medbreak
\textbf{Example 3 (discretized Bertrand with logistic demand)} Consider a symmetric Bertrand duopoly in which two firms sell a differentiated product at marginal cost $c$. Assume the following standard demand function for firm $A$ under the profile of prices $(p_A,p_B)$
\begin{equation}D_A(p_A,p_B)=\frac{e^{-\frac{a-p_A}{\lambda}}}{1+ e^{-\frac{a-p_A}{\lambda}} + e^{-\frac{a-p_B}{\lambda}} }
\end{equation}
where $\lambda>0$ and $a>0$ control for the degree of differentiation. The profit of firm $A$ under profile $(p_A,p_B)$  is
\begin{equation}
u(p_A,p_B)=D_A(p_A,p_B)(p_A-c)
\end{equation}
Denote $p_N$ the (unique and symmetric) Nash equilibrium price and $p_M$ the monopoly price for this duopoly. Assume the algorithms can select prices on a grid of prices $A=\{p_1...p_K\}$ above $p_N$ and below $p_M$. The pricing game in which players can select prices from $A$ satisfies the assumptions $(1), (2)$ and $(3)$ (see Appendix \ref{appendix: Bertrand}).
\subsection{Q-learning}
Players delegate their decision making to Q-learning algorithms. Q-learning is a simple reinforcement learning principle designed to find optimal solutions to optimization problems in a Markov environment. In this paper, the \textit{memory-less} version is considered. A more detailed introduction to Q-learning is given in Appendix \ref{Appendix: q-learning}.\medbreak For each action $a\in \mathcal{A}$, the algorithm keeps a \textit{Q-value} in memory and updates it as play proceeds. The Q-value of action $a$ at time $t$ is denoted $Q_t(a)$. The algorithm performs two tasks. At any time $t$, it selects an action using an \textit{exploration policy}. Once the actions have been played and the payoffs have been received, it updates its Q-values using the \textit{update rule}.
\subsubsection*{Update rule}
Denote $a'_t$ the action played by the opponent at time $t$. The update rule used by Q-learning is, for all $a\in \mathcal{A}$:
\begin{equation} \label{eq: update}
\left\{
    \begin{array}{ll}
    Q_{t+1}(a)=(1-\alpha) Q_t(a) + \alpha \Big{[} u(a,a_t') + \gamma \max_{a''\in A} Q_t(a'') \Big{]} \text{ if $a$ is played at $t$}
    \\
    Q_{t+1}(a)=Q_{t}(a) \text{ otherwise}
    \end{array}
\right.
\end{equation}
where $\alpha\in [0,1]$ is referred to as the \textit{learning rate} and controls how rapidly Q-values change when a payoff is observed. When $\alpha=0$ the Q-values stay constant, so that the agent does not learn, and when $\alpha=1$ the Q-values change immediately. The parameter $\gamma$ can be classically seen as a discount factor. In this paper, players use the limit of the mean criterion, so that $\gamma$ is rather viewed as a learning parameter, which, like $\alpha$, is used to tune the algorithm.\footnote{In some articles, for instance Colliard et al. (2023) \cite{colliard2022algorithmic}, $\gamma$ is exogenously set to $0$. Such an assumption does not change any of the results in the present paper.}
\subsubsection*{Exploration policy}
To decide which action to take given a vector of Q-values, the agent relies on an \textit{exploration policy}. We assume that the players use the standard $\varepsilon$-greedy policy. Under $\varepsilon$-greedy, the action chosen at time $t$ is
\begin{equation}
    a_{t}= \left\{
    \begin{array}{ll}
    \arg \max_{a'} Q_{t}(a') \text{ with probability } 1-\varepsilon
          \\
    \sim \mathcal{U}(\mathcal{A}) \text{ with probability } \varepsilon.
    \end{array}
\right.
\end{equation}
The parameter $\varepsilon$ trades-off \textit{exploration} and \textit{exploitation}. The higher $\varepsilon$, the more frequently the algorithm explores. For $\varepsilon=1$ the algorithm always chooses uniformly at random. Conversely, when $\varepsilon=0$ the algorithm \textit{is greedy}: it always chooses the action with highest Q-value.
\subsubsection*{The greedy policy}
Under the \textit{greedy} policy, the algorithm only selects actions with highest Q-value. If more than one action has highest Q-value, then we will assume that it selects one of these actions randomly and uniformly. The update rule for the greedy policy takes a simpler form allowing to characterize the variations of the Q-values. It becomes:
\begin{equation}
\left\{
    \begin{array}{ll}
    Q_{t+1}(a)=(1-\alpha) Q_t(a) + \alpha \Big{[} u(a,a_t') + \gamma Q_t(a) \Big{]} \text{ if $a$ is played at $t$}
    \\
    Q_{t+1}(a)=Q_{t}(a) \text{ otherwise}
    \end{array}
\right.
\end{equation}
Thus, the Q-value of action $a$ increases at $t$ if and only if
\begin{equation}
    Q_t(a)\le \frac{u(a, a_t')}{1-\gamma}.
\end{equation}
Whether or not $Q_t(a)$ increases only depends on the action played by the opponent and the current Q-value. The quantity $\frac{u(a, a_t')}{1-\gamma}$ is the $\gamma$-discounted payoff collected by the algorithm when it plays action $a$ and the opponent plays action $a_t'$ for ever. If $Q_t(a)$ increases at $t$ then the action $a$ is reinforced: it will be played next period by the greedy algorithm, since the order of its Q-values remains unchanged. Note also that, for any policy, the Q-values can never grow unbounded: whenever the action with maximal Q-value is updated, it decreases if it lies above $\max_{a'\in A}\frac{u(a, a')}{1-\gamma}$.
\subsubsection*{Initial conditions}
Q-learning algorithms requires that an initial vector of Q-values be chosen prior to letting them operate. We shall assume that for each action and player, the initial corresponding Q-value is chosen from the interval $I_a=[\frac{\min_{a,a'}u(a,a')}{1-\gamma}, \max_{a,a'}u(a,a') + \frac{\gamma}{1-\gamma} \max_{a', a''} u(a',a'')]$, the bounds of which correspond to the lowest and highest limiting values the Q-values can take. The two vectors of Q-values are drawn according to a distribution $\mathcal{Q} \in \Delta\big(\big(\bigtimes_{a \in \mathcal{A}} I_a \big)^2\big)$, which we assume has full support. We interpret this assumption by the presence of a first initial period of exploration (like in Lambin 2024 \cite{lambin2024less}) by the Q-learning algorithms prior to behaving according to prescribed $\varepsilon$-greedy rule. The following result justifies the choice of this interval for the initial Q-values.
\begin{proposition} \label{prop: natural interval}
    Assume both players select all actions with positive probability bounded away from zero. Take two arbitrary vectors of initial Q-values $Q_0^A$ and $Q_0^B$ in $\mathbb{R}^n$. Then with probability $1$, there exists $T$ such that for all $t>T$, for both algorithms and for any action $a$, $Q_t(a)\in I_a$.
\end{proposition}
This result relies on the fact that, whenever Q-values lie above the upper bound of this interval, they decrease when they are updated. When they reach the interior of this interval, they remain inside. Conversely, if Q-values are below the lower bound of this interval, they keep increasing whenever they are updated. Since both players use a non-greedy policy, with probability $1$, all Q-values are updated enough times to get inside their $I_a$. In Appendix \ref{Appendix: different intialization}, we consider a limiting case in which the Q-values are initialized as the average of the payoffs given by their actions against the others and show that our results carry on with minimal adjustments. We also consider the case in which initial conditions can be chosen by $A$ and $B$, and show that our results are not modified.
\subsubsection*{Payoffs to the players} \label{Sec: payoffs}
Denote $a_t$ the action played by $A$'s algorithm at time $t$ and $a_t'$ the action played by $B$'s algorithm at time $t$. $A$ and $B$ evaluate their payoffs under the profile $(\varepsilon_A, \varepsilon_B)$ using the limit of the mean criterion. Denoting $\Pi_A(\varepsilon_A, \varepsilon_B)$ the expected payoff of player $A$ under the profile $(\varepsilon_A, \varepsilon_B)$:
\begin{equation}
    \Pi_A(\varepsilon_A, \varepsilon_B) = \mathbb{E}_{\varepsilon_A, \varepsilon_B}\Big[ \liminf _{T \rightarrow +\infty} \frac{1}{T}\sum_{t=1}^{T}u(a_t,a_t')\Big]
\end{equation}
Similarly, the expected payoff of player $B$ under the same profile is denoted $\Pi_B(\varepsilon_A, \varepsilon_B)$ and writes 
\begin{equation}
    \Pi_B(\varepsilon_A, \varepsilon_B) = \mathbb{E}_{\varepsilon_A, \varepsilon_B}\Big[ \liminf _{T \rightarrow +\infty} \frac{1}{T}\sum_{t=1}^{T}u(a_t',a_t)\Big]
\end{equation}
\subsection{The two-stage game}
The game played by $A$ and $B$ is denoted $G(G_0)$, where $G_0$ is a game belonging to $\mathcal{G}$. Formally
\begin{equation}
    G(G_0)=<\{A,B\}, \ [0,1]^2, \ \Pi=(\Pi_A, \Pi_B) >
\end{equation}
This game is a \textit{supergame} of $G_0$ with strategies restricted to Q-learning algorithms with $\varepsilon$-greedy exploration policies whose parameters are chosen in $[0,1]$.
\begin{definition}
    A profile of strategies $(\varepsilon_A, \varepsilon_B)$ is said to be \textit{collusive} if and only if for all $X \in \{A,B\}$,
    \begin{equation}
        \Pi_X(\varepsilon_A,\varepsilon_B)>u(a_1,a_1).
    \end{equation}
\end{definition}
Under a collusive profile, both players receive a higher payoff than they would under the strict Nash equilibrium of the one-shot game $G_0$. The difference between \textit{collusive} and \textit{individually rational} lies in the inequality being strict. The remainder of this article aims to answer the following question: are Nash equilibria of $G(G_0)$ collusive? \medbreak In the two-stage game, the margin of competition moves from the actions of $G_0$ to the selection of exploration policies. If there is an exploration policy allowing to play non-cooperatively while the opponent does, then one could expect algorithmic collusion to be prevented. The answer to the previous question is yes, all (pure) Nash equilibria of the two-stage game are collusive.
Since collusion results from the strategic ex-ante selection of learning algorithms, an effective competition policy must target the selection stage itself—by regulating or constraining the choice of exploration parameters. The next section is devoted to proving this result formally.
\section{Collusion in equilibrium} \label{sec: collusion in eq}
\begin{theorem} \label{Thm1}
    Let $G_0\in \mathcal{G}$. If $(\varepsilon_A,\varepsilon_B)$ is a Nash equilibrium of $G(G_0)$, then $(\varepsilon_A, \varepsilon_B)$ is collusive.
\end{theorem}
As an intermediary result to prove this theorem, I will also prove that the set of Nash equilibria is not empty. The theorem relies on players being allowed to choose greedy algorithms. The greedy exploration policy has \textit{limited forgiveness}\footnote{This terminology is purposely close to the one of Axelrod (1980) \cite{axelrod1980effective}, who highlights the success of \textit{forgiving} strategies in the repeated prisoner's dilemma. }. It leaves scope for mutual cooperation, but reacts to abusively uncooperative behavior by settling on the least cooperative action. As will be shown in the section dedicated to numerical simulations, this intuition remains valid for positive exploration parameters. \medbreak 
As a first step into the proof of Theorem \ref{Thm1}, one needs the following simple lemma.
\begin{lemma} \label{L1}
    Assume player $A$'s algorithm plays a constant action $a\in \mathcal{A}$ at all times. Denote $BR(a)$ the set of best-responses to action $a$. Assume player $B$ chooses $\varepsilon_B>0$. Then for all $a^* \in BR(a)$, there exists $t^*$ such that for all $a' \notin BR(a)$: 
    \begin{equation}
         \forall t\ge t^*, Q_t^B(a^*)\ge Q_t^B(a').
    \end{equation}
\end{lemma}
Lemma \ref{L1} states that a non greedy algorithm facing a constant action eventually \textit{learns} a best response to this action. Here learning has a specific meaning. By "learning" it is meant that the algorithm ends up giving a higher value to an action that performs best against $a$. However it does not mean that the algorithm only plays this action. To the contrary, since by assumption $B$ uses a non-greedy policy, it explores with non-zero probability and thus plays non best-response actions with positive probability. It is only to this price that the algorithm is able to adapt to its opponent's action: randomly sampling actions allows one to eventually detect the ones that performs best. Lemma \ref{L1} is close to the classical convergence guarantees given by Sutton and Barto (1998) \cite{sutton2018reinforcement}, to the difference that it does not assume the usual conditions on the exploration and learning rates (see Appendix \ref{Appendix: q-learning}). \medbreak A Q-learning algorithm with a non-greedy policy facing a constant action thus leaves no profit opportunity. In the prisoner's dilemma, this means that a naively cooperative player will be detected and exploited by the non-greedy Q-learning algorithm. This can be viewed as an advantage of Q-learning over more classical strategies in repeated games, like tit-for-tat. Indeed, like tit-for-tat, Q-learning algorithms are able to sustain some cooperation, unlike tit-for-tat however, they are not overly generous with naive players.\medbreak
The next proposition characterizes the limiting behavior of the algorithms when one player ($A$) chooses $\varepsilon_A=0$ and the other ($B$) chooses $\varepsilon_B>0$.
\begin{proposition}\label{prop: asymmetric exploration}
For all $\varepsilon_B>0$, $\Pi_A(0, \varepsilon_B)>u(a_1,a_1)$ and $\Pi_B(0, \varepsilon_B)<u(a_1,a_1)$.
\end{proposition}
In this case, the player that chose the greedy algorithm is better-off compared to the one who chose a positive level of exploration. This is due to the limiting behavior of the algorithms: in this case both algorithms eventually learn the least cooperative action $a_1$. However, since $A$ chose the greedy exploration policy, with probability $1$, his algorithm will settle to play $a_1$ for all $t$ large enough. On the other hand, $B$ chose a non-greedy exploration policy. As a consequence, his algorithm plays other actions with positive probability. The payoff of player $B$ writes
\begin{equation} \label{Eq payoff non greedy}
    \Pi_B(\varepsilon_A,\varepsilon_B)=(1-\frac{K-1}{K}\varepsilon_B)u(a_1,a_1)+ \sum_{a\ne a_1} \frac{K-1}{K} \varepsilon_B u(a, a_1)
\end{equation}
while this of $A$ writes
\begin{equation}
    \Pi_A(\varepsilon_A,\varepsilon_B)=\sum_{a \in A}\frac{K-1}{K}\varepsilon_Bu(a_1, a)
\end{equation}
Since $(a_1,a_1)$ is a strict Nash equilibrium in $G_0$, each time $B$'s algorithm explores and selects $a\ne a_1$, it collects a payoff strictly lower than $u(a_1,a_1)$. To the contrary, since in $G_0$, a player is better-off when his opponent plays more cooperatively each time $B$ explores, $A$'s algorithm collects a payoff higher than $u(a_1,a_1)$. The greedy exploration policy allows one to secure more than the incentive compatible payoff. This element is one of the key argument in establishing Theorem \ref{Thm1}. The proof of this lemma relies on two steps. First, we show that $A$'s algorithm ends up holding $a_1$ as a preferred action with probability $1$. For any action $a$ played by $A$'s algorithm, whenever $B$'s algorithm plays $a_1$ the Q-value of $A$'s algorithm for action $a$ necessarily decreases. If a long enough sequence of action $a_1$ is played by $B$'s algorithm, all the other actions' Q-values of $A$'s algorithm decrease and eventually become lower than $\frac{u(a_1,a_1)}{1-\gamma}$. Since $B$'s algorithm explores with positive probability, this happens with probability $1$. When this happens, the Q-value of $a_1$ is irreversibly above the others and $A$ then plays $a_1$ for ever on. Second, we use Lemma \ref{L1} to show that, facing the constant action $a_1$, $B$'s algorithm adapts and learns a best-response, that is $a_1$. \medbreak
The greedy exploration policy's tolerance is limited: as long enough sequences of uncooperative behavior by the opponent happen, it ends up playing uncooperatively and cannot switch back to cooperative behavior. When the behavior of player $A$'s algorithm becomes constant, $B$'s algorithm can eventually adapt and learn to play uncooperatively. \medbreak The final lemma states that when both players choose a greedy algorithm ($\varepsilon_A=\varepsilon_B=0$) then their payoffs are strictly above $u(a_1,a_1)$.
\begin{proposition}\label{prop: 00}
\begin{equation}
\Pi(0,0)>u(a_1,a_1)
\end{equation}
\end{proposition}
In this case, only the maximal Q-values are updated for both algorithms. The evolution of the Q-values is thus fully deterministic. Since only the maximal Q-values are updated, all the other Q-values are constant. As a consequence, for both algorithms the second highest Q-value is decreasing. Indeed, the second highest Q-value can only change when there is a change in the order of the Q-values, which only happens when the highest Q-value undercuts the second highest. The proof of Proposition \ref{prop: 00} then proceeds in two steps. First, we show that for both players, in any realization of the initial conditions, the limit inferior of the mean payoff is necessarily (weakly) higher than $u(a_1,a_1)$. This is done by distinguishing two cases according to the limit of the second highest Q-value. If it is strictly lower than $\frac{u(a_1,a_1)}{1-\gamma}$ then the Q-value for $a_1$ must remain above all others starting at some time $t$. In this case, both algorithms end up playing $a_1$ for ever on, and they thus collect $u(a_1,a_1)$. If it is higher than $\frac{u(a_1,a_1)}{1-\gamma}$, we show that it cannot be the case that the algorithm receives a sequence of payoffs whose limit inferior is smaller than $\frac{u(a_1,a_1)}{1-\gamma}$. Second, we make the point that, for some initial conditions, both algorithms reinforce themselves into cooperative behavior, in which case they collect $u(a_K,a_K) > u(a_1,a_1)$. Indeed, since the algorithms are both greedy, and since by assumption a player is always better off in $G_0$ when his opponent plays more cooperatively, when both algorithms select $a_K$, then the associated Q-values increase for both. In this case, the order of Q-values is unchanged, and the algorithms will play the same profile next period. This proves that $(0,0)$ is a collusive profile. \medbreak Proposition \ref{prop: asymmetric exploration} and \ref{prop: 00} yield the following corollary.
\begin{corollary} 
\label{cor: greedy greedy € NE}
    The profile $(\text{greedy},\text{greedy})$ is a strict (collusive) Nash equilibrium.
\end{corollary}
\begin{proof}
   We shall check that there is no profitable deviation. Assume $\varepsilon_B=0$. Under the greedy exploration policy, by Proposition \ref{prop: 00}, the payoff to player $A$ is greater than $u(a_1,a_1)$. If player $A$ selects an exploration policy $\varepsilon_A>0$, then by Proposition \ref{prop: asymmetric exploration}, he gets a payoff strictly lower than $u(a_1,a_1)$. 
\end{proof}
The limited forgiveness of the greedy exploration policy ensures that the opponent has no incentive to choose a non-greedy policy. It also leaves enough scope for mutual cooperation when the opponent chooses a greedy policy. It thus allows for a form of \textit{reactive commitment}: cooperation is possible, but attempts to abuse it are punished. Tit-for-tat uses a similar principle to sustain cooperation in the repeated prisoner's dilemma. The important difference is that, unlike tit-for-tat, a greedy Q-learning algorithm needs no prior bias towards cooperation. \medbreak The previous proposition guarantees that the set of Nash equilibria is not empty so that Theorem \ref{Thm1} is meaningful. Theorem \ref{Thm1} is obtained by combining the previous results.\medbreak \textbf{Proof of Theorem \ref{Thm1}} Assume, by way of contradiction, that $(\varepsilon_A, \varepsilon_B)$ is a non-collusive Nash equilibrium. Since $(\text{greedy},\text{greedy})$ is collusive (Corollary \ref{cor: greedy greedy € NE}), then either $\varepsilon_A\ne 0$ or $\varepsilon_B\ne 0$. Further, since $(0,0)$ is a strict Nash equilibrium, the unique best response to $0$ is $0$ so that necessarily $\varepsilon_A \ne 0$ and $\varepsilon_B \ne 0$. By assumption, either $\Pi_A(\varepsilon_A,\varepsilon_B)\le u(a_1,a_1)$ or $\Pi_B(\varepsilon_A,\varepsilon_B)\le u(a_1,a_1)$. Assume without loss of generality that $\Pi_A(\varepsilon_A,\varepsilon_B)\le u(a_1,a_1)$. Then by Proposition \ref{prop: asymmetric exploration} a profitable deviation to the greedy exploration policy exists for player $A$. This contradicts $(\varepsilon_A, \varepsilon_B)$ being an equilibrium. \qedsymbol \medbreak Both players can ensure a strictly supra-competitive payoff by using the greedy exploration policy. The immediate consequence is that any equilibrium is collusive. This result has an important implication for algorithmic collusion. The legitimate concerns raised by tacit collusion of reinforcement learning algorithms cannot be alleviated by letting the players freely choose their exploration policies. Furthermore, Q-learning algorithms require no bias towards supra-competitive actions. Regulation should therefore be directed at the reinforcement learning technology itself for algorithmic collusion to be prevented.

\section{Numerical experiments} \label{sec: simulations}
In this section, we run extensive numerical simulations to computationally find Nash equilibria and uncover the role of the exploration levels. We first consider a duopoly \textit{à la} Bertrand. Our simulations reveal that the agents tend to \textit{over-explore} in equilibrium, to the detriment of joint profit. We then restrict attention to the simpler case of the parameterized Prisoner's dilemma\footnote{The parameterization used is the same as the one studied by Banchio and Mantegazza 2023} to analyze in detail the strategic role of exploration levels. Two main tasks are performed: (i) the intensity of collusion is measured and (ii) pure strategy Nash equilibria of $G$ are computationally identified. Their results highlight that players have an incentive to raise their own exploration policy so as to try and exploit the other algorithm's cooperation. However, if a player selects a level of exploration that is too high, his opponent's algorithm reacts by turning uncooperative. We finally consider an extension of the parameterized prisoner's dilemma with $N\ge 2$ players and show how increasing the number of players decreases the level of payoffs to bring it closer to the Nash equilibrium one. 
\subsection{Bertrand duopoly}
In this section, we consider a discretized Bertrand duopoly with logistic demand, as described in example $3$ of Section \ref{sec: One-shot game}, in which the players' algorithms can select prices from a grid of $n=15$ prices between the competitive (Nash) price and the monopoly price. We choose a parameterization close to that of Calvano et al. (2020) and of Lambin (2024) and set $\lambda=0.25$, $a=2$ and $c=1$. We run the Q-learning algorithms with $K \times K = 40 \times 40$ pairs of distinct values for $\varepsilon_A$ and $\varepsilon_B$ equally spaced in $[0,1]$. For each couple of exploration parameters, we run the algorithms $M=50$ times over $T=2\times 10^6$ time periods and obtain the limiting payoffs (the average payoff obtained over the last $10^3$ time periods). This way, we compute two payoff matrices and obtain the corresponding pure strategy Nash equilibria. 
\begin{figure} 
    \centering
    \includegraphics[width=0.8\linewidth]{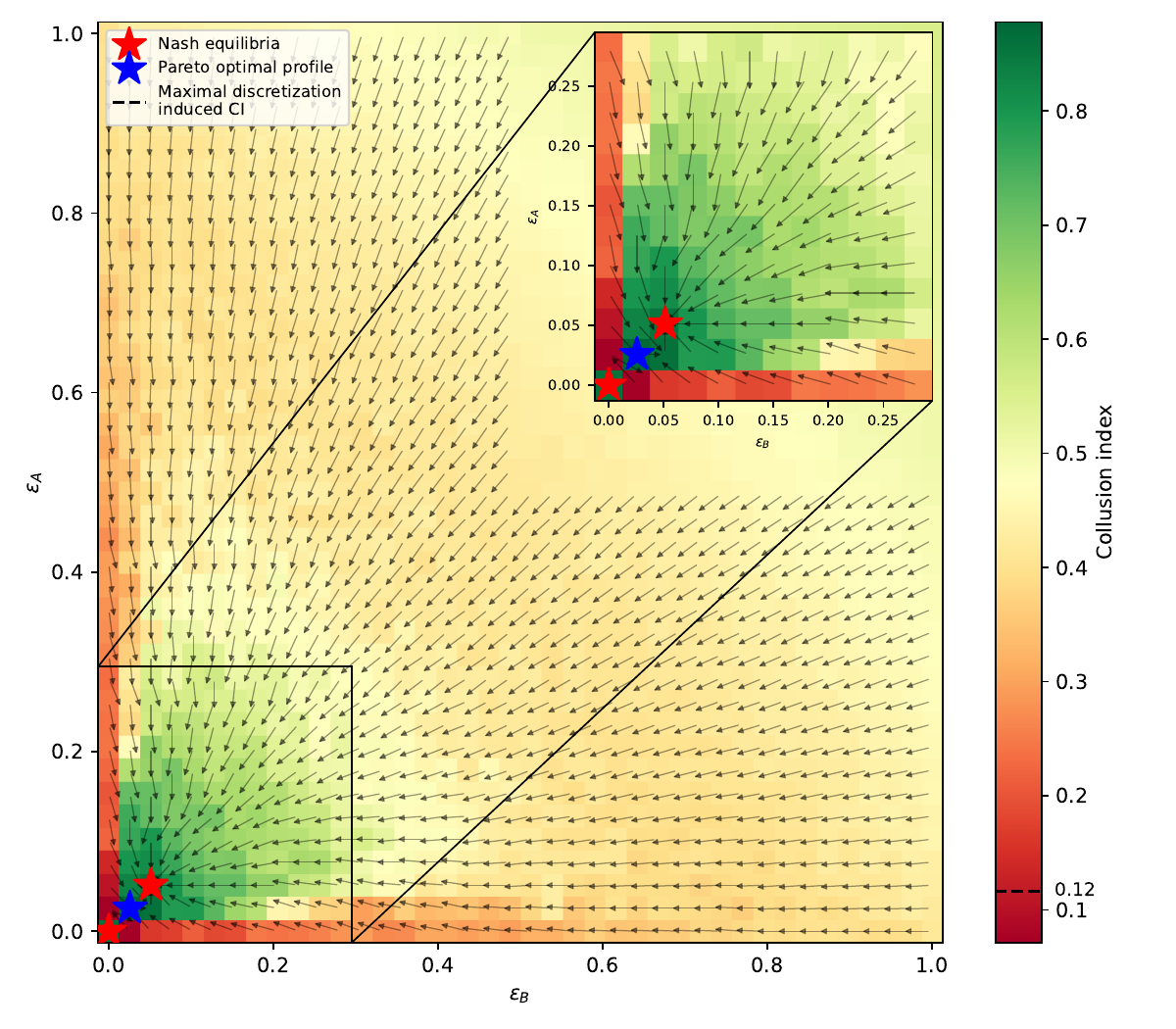}
    \caption{Results of the numerical experiments for the Bertrand duopoly}
    \label{fig: results_duopoly}
\end{figure}
Figure \ref{fig: results_duopoly} displays the results of the numerical experiments. For each couple of parameters $(\varepsilon_A, \varepsilon_B)$ we compute a standard collusion index by the formula:
\begin{equation}
    \text{CI} = \frac{\Pi_A(\varepsilon_A, \varepsilon_B) + \Pi_B(\varepsilon_A, \varepsilon_B) - 2u(a_1,a_1)}{2u(a_n, a_n) - 2u(a_1,a_1)}
\end{equation}
where $u(a_1,a_1)$ and $u(a_n,a_n)$ are respectively the payoffs obtained by a player in the Nash equilibrium of the Bertrand duopoly and under monopoly pricing. The collusion index of each pair of exploration parameters is represented by the color of the corresponding cell, the brighter the more collusion there is. We note that some collusion appears, as the recorded levels of the collusion index can be well above the collusion index induced by the discretization of the price space (indicated by a dashed line on the color scale). There is more collusion for moderate values of exploration, which is in line with the theoretical results of Banchio and Mantegazza (2022). Figure \ref{fig: results_duopoly} also shows that no collusion appears when one of the algorithms is greedy while the other is not, as we have proved in Proposition \ref{prop: asymmetric exploration}. Notable profiles of exploration are indicated with colored stars: red for Nash equilibria, blue for the profile maximizing the joint profit. As we have proved in Corollary \ref{cor: greedy greedy € NE}, $(0,0)$ is a Nash equilibrium. Another Nash equilibrium exists, it is symmetric and has a higher level of exploration than the profile that maximizes joint profit. There is thus an equilibrium with \textit{over-exploration}\footnote{When we look for $\eta$-equilibria (for $\eta$ small) instead of pure strategy Nash equilibria, a similar pattern happens: symmetric $\eta$-equilibria with over-exploration appear.}. On each cell of Figure \ref{fig: results_duopoly} we represent an arrow indicating the direction of the best-response. One can see that the field points towards the over-exploration equilibrium rather than towards the $(0,0)$ one. In the following, we will thus focus attention on the equilibria that are not the $(0,0)$ one. We argue that the tendency to over-explore in equilibrium is due to the specific strategic role of exploration levels, which we uncover by devoting the next section to an extensive numerical analysis in the case of the prisoner's dilemma.
\subsection{The parameterized Prisoner's dilemma}
In this section we focus attention to a parameterized prisoner's dilemma like in Banchio and Mantegazza (2022). Focusing on this simple special case will allow to precisely measure the effect of exploration levels on the behavior of the algorithms as well as to assess how they can be used by players to unilaterally increase their payoffs and how this comes at the cost of decreasing the joint payoff. Finally, we will consider an $N$ players extension and show that this effect is responsible for an even stronger decrease in joint payoff.
\subsubsection{Preliminaries}
\inlinetitle{The one-shot game}
The payoffs of the one-shot game, denoted $G_0(g)$, are given in Table \ref{payoffs}. The parameter $g$ takes values in $(1,2)$ and is interpreted as the value of cooperation. It has two effects on the game.
First, since $u(C,C)-u(D,D)=2(g-1)$, when $g$ increases mutual defection becomes more socially detrimental. In the limit case $g=1$, $u(C,C)=u(D,D)$ so that there is no social cost to mutual defection. Second, as $u(D,C)-u(C,C)=2-g$ is decreasing with $g$, the incentive to deviate from mutual cooperation is decreasing with $g$ and vanishes as $g$ approaches $2$. $G_0(g)$ can be interpreted as a contribution game in which two players begin with an endowment of two monetary units and need to choose whether to invest it in a common pool. The common pool then grows by a factor $g$ and is equally split between the two players. If a player does not invest their endowment in the pool, they get to keep it yet still receive half of the pool after it has grown.
\FloatBarrier
\begin{table}
\centering
    \setlength{\extrarowheight}{2pt}
    \begin{tabular}{*{4}{c|}}
      \multicolumn{2}{c}{} & \multicolumn{2}{c}{Player $B$}\\\cline{3-4}
      \multicolumn{1}{c}{} &  & $C$  & $D$ \\\cline{2-4}
      \multirow{2}*{Player $A$}  & $C$ & $2g,2g$ & $g,2+g$ \\\cline{2-4}
      & $D$ & $2+g,g$ & $2,2$ \\\cline{2-4}
    \end{tabular}
    \caption{Payoffs in the parameterized Prisoner's Dilemma}
    \label{payoffs}
  \end{table} \medbreak
\inlinetitle{Spontaneous coupling}
This problem has been studied by Banchio and Mantegazza (2023) in the special case $\varepsilon_A=\varepsilon_B=\varepsilon$. Using continuous time equivalents, they prove that for moderate values of $\varepsilon$, the player's algorithms reach \textit{spontaneous coupling}. Under spontaneous coupling, phases of cooperation alternate with phases of defection, leading to a collusive outcome. The space of Q-values is divided into different regions. For $P\in \{A,B\}$, $X\in \{C,D\}$, let $\omega_{P}^X=\{\mathbb{Q} = (Q_A^C, Q_A^D, Q_B^C,Q_B^D)\in \mathbb{R^4} \text{ s.t. } Q_P^X>Q_{P}^{-X}\}$ where $-X=C$ when $X=D$ and conversely. Further, for $X,Y \in \{C,D\}$ let $\omega_{X,Y}=\omega_{A}^X\cap \omega_{B}^Y$. Geometrically, $\omega_P^X$ corresponds to the half $\mathbb{R}^4$ space in which algorithm $P$ holds action $X$ as a preferred action. Similarly, $\omega_{X,Y}$ corresponds to the quarter of space in which algorithms $A$ and $B$ respectively hold actions $X$ and $Y$ as preferred actions. The average time spent, in the limit, in the region $\omega_{X,Y}$ is denoted $\tau_{X,Y}$, and is a function of $\varepsilon_A, \varepsilon_B$ and $g$. Banchio and Mantegazza (2023) are able to give a closed-form expression of $\tau_{X,Y}$ under $\varepsilon_A=\varepsilon_B=\varepsilon$. It reaches $0$ when $\varepsilon$ is larger than a threshold $\bar{\varepsilon}(g)$, which increases with $g$. However their proof heavily relies on the symmetry assumption and cannot be extended to general profiles with $\varepsilon_A\ne \varepsilon_B$. Here spontaneous coupling for general profiles is investigated using numerical simulations. \medbreak
\inlinetitle{The effect of $\varepsilon$ on individual payoffs}
One's level of exploration, say $\varepsilon_A$, has an effect over $A$'s payoffs through two channels. First, increasing $\varepsilon_A$ changes the amount of time spent in each of the four regions $\omega_{XY}$. Second, it has a direct effect on $A$'s payoffs by affecting the probability to play the action with the lowest Q-value. The latter effect depends on the time spent in the different zones. When $A$ holds $C$ as a preferred action, an increase in $\varepsilon_A$ has a positive effect on $A$'s payoffs. To the contrary, when $A$ holds $D$ as a preferred action, an increase in $\varepsilon_A$ has a negative effect on his payoffs. As long as an increase in $\varepsilon_A$ does not decrease too much the time during which he holds $C$ as a preferred action, increasing $\varepsilon_A$ has a positive direct payoff effect for $A$. To the contrary, if increasing $\varepsilon_A$ leads to preventing spontaneous coupling to appear, it is detrimental to player $A$. This effect is one of the important ingredients that characterize the strategic role of exploration levels. The others are related to time spent in each zones, and to the appearance or not of spontaneous coupling. They are investigated in the following section.
\subsubsection{Measurements of spontaneous coupling}
\begin{figure}[ht]
\centering
    \includegraphics[width=0.75\textwidth]{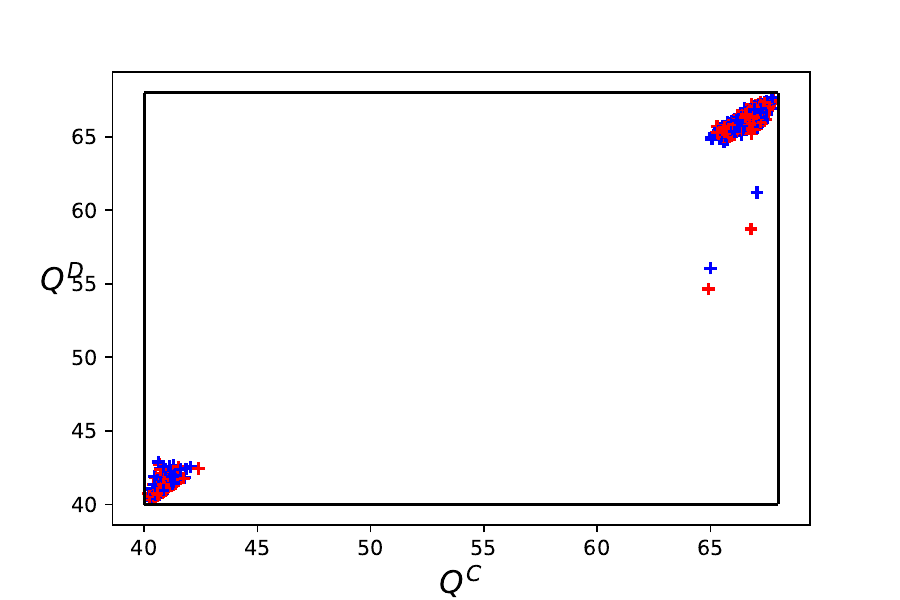}
    \caption{\textbf{A visual representation of spontaneous coupling}}
    \label{fig: coupling}
\end{figure}

\begin{figure}[ht]
        \begin{minipage}[b]{0.32\linewidth}
            \centering
            \includegraphics[width=\textwidth]{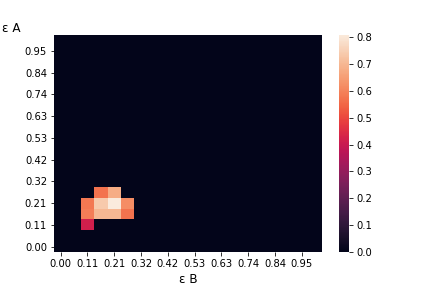}
            \caption*{$g=1.42$}
        \end{minipage}
        \begin{minipage}[b]{0.32\linewidth}
            \centering
        \includegraphics[width=\textwidth]{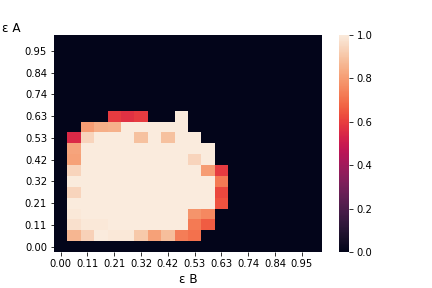}
            \caption*{$g=1.68$}
        \end{minipage}
        \begin{minipage}[b]{0.32\linewidth}
            \centering
        \includegraphics[width=\textwidth]{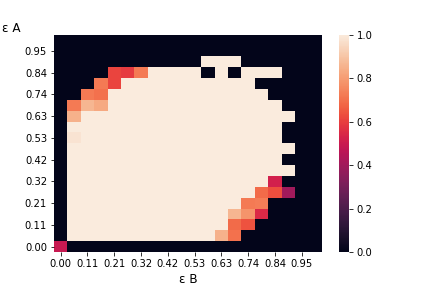}
            \caption*{$g=1.84$}
        \end{minipage}
    \caption{\textbf{Size of the spontaneous coupling cluster}}
    \label{fig: bassin_measured}
\end{figure}

\begin{figure}[ht]
        \begin{minipage}[b]{0.32\linewidth}
            \centering
            \includegraphics[width=\textwidth]{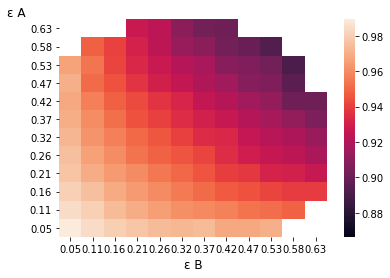}
            \caption*{$\omega_{CC} \rightarrow \omega_{CC}$}
        \end{minipage}
        \begin{minipage}[b]{0.32\linewidth}
            \centering
        \includegraphics[width=\textwidth]{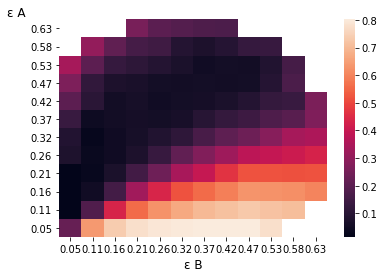}
            \caption*{$\omega_{CD} \rightarrow \omega_{CD}$}
        \end{minipage}
        \begin{minipage}[b]{0.32\linewidth}
            \centering
        \includegraphics[width=\textwidth]{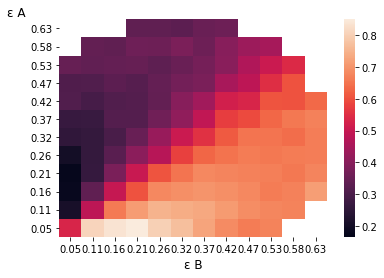}
        \captionsetup{font=footnotesize}
            \caption*{$\omega_{CC} \rightarrow \omega_{CD} | \omega_{CC} \rightarrow \omega_{CD}\cup \omega_{DC}$}
        \end{minipage}
    \caption{\textbf{Transitions probabilities for $g=1.7$}}
    \label{fig: transition}
    \end{figure}

\begin{figure}[ht]
\centering
    \includegraphics[width=0.7\textwidth]{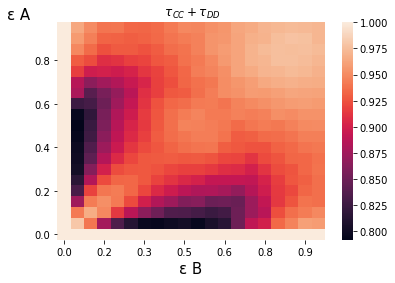}
    \caption{\textbf{Average fraction of time spent in a symmetric zone}}
    \label{fig: symmetric}
\end{figure}

\inlinetitle{Methodology}
The values of $\gamma$ and $\alpha$ are fixed throughout this section to $0.95$ and $0.1$ respectively. The other parameters, $g\in (1,2), \varepsilon_A, \varepsilon_B \in [0,1]$ vary and are selected on an equally spaced grid with $20^3$ triplets. For each triplet $(g,\varepsilon_A,\varepsilon_B)$, $G_0(g)$ is generated and the process is simulated between $50$ and $100$ times. Metrics of interest (payoff, position of Q-values etc.) are averaged over the last $100$ to $1000$ periods of the process and over the $50$ to $100$ experiments. \medbreak 
\inlinetitle{Detecting and measuring spontaneous coupling} \label{sec: bassin}
Spontaneous coupling can be schematically explained as follows. In $\omega_{CC}$ phases, the Q-values for both $C$ and $D$ increase, however Q-values for $D$ tend to increase faster causing a switch back to $\omega_{DD}$. In $\omega_{DD}$ phases, Q-values for both $C$ and $D$ tend to decrease. This leaves a possibility to transition from $\omega_{DD}$ back to $\omega_{CC}$. This alternating behavior can only appear when the exploration policies are low enough. Indeed, the Q-values for $C$ should not be updated too frequently so as not to decrease faster than the ones for $D$. Consequently, the Q-values for both actions tend to be higher when the algorithms reach spontaneous coupling compared to when they do not. This can be visualized on Figure \ref{fig: coupling}.
Figure \ref{fig: coupling} displays the position of $k=1000$ independent runs of the algorithm's Q-values at $t=10^5$, with $g=1.7$ and $\varepsilon_A=\varepsilon_B=0.1$. They are represented as dots of different colors: blue for player $A$ and red for player $B$. Two clusters appear, one on the top right corner (high Q-values), one on the bottom left corner (low Q-values). The first one corresponds to spontaneous coupling being reached, the second to the other case. The first task is to detect and measure the sizes of these clusters, so as to have a sense of how collusive is a triplet of parameters $(g, \varepsilon_A, \varepsilon_B)$. For this purpose, two distinct clustering tasks are performed using a K-means algorithm \cite{lloyd1982least, forgy1965cluster} with two clusters (K=2). \medbreak K-means is a well-known algorithm used to perform unsupervised learning. Given a set of points in a space, it aims at assigning \textit{observations}, points in a metric space, to clusters, which form a partition of the set of observations. The quality of a clustering is measured with an inertia metric. Formally, given $(E,d)$ a metric vector space, $X=\{X_i,i\in[1,...,n]\}$ a set of $n$ observations, and $C=\{C_1...C_K\}$ a partition of $X$, the inertia metric is given by
\begin{equation}
    I(C,X)=\sum_{i=1}^K\sum_{k=1}^{|C_i|}d(X_k,\mu_i)
\end{equation}
where $\mu_i=\frac{1}{|C_i|}\sum_{k\in C_i}X_k$ is the \textit{barycenter} of cluster $C_i$. Even though K-means is known to perform well for simple tasks, it is a heuristic method that does not necessarily produce the best clustering. Given the simplicity of the task, a more advanced method is not necessary. The algorithm takes as input a list of observations, in this case a list of $4$-dimensional points, and a number of clusters (here $K=2$), and outputs the centers of the identified clusters as well as the cluster to which each observation belongs, which enables to compute the inertia metric of the achieved clustering. It proceeds as follows: first, it chooses randomly $K$ observations as clusters' centers and assigns each observation to the cluster with nearest center. Then, it updates the centers of the clusters given the new partition of observations and repeats the operation. The process stops whenever the clusters' centers do not move anymore. \medbreak In the first task, $\tau_{CC}$ is measured by counting the number of periods the algorithms spend in $\omega_{CC}$ out of the last $1000$ periods. Then triplets of parameters are classified into two categories, either \textit{allowing for spontaneous coupling} or \textit{not allowing for spontaneous coupling}, using $\tau_{CC}$ as a feature. The algorithm successfully identifies one cluster with low values of $\tau_{CC}$ and another with higher values, the second being the triplets of parameters allowing for spontaneous coupling. Then, for the triplets of parameters for which spontaneous coupling is detected, $k=1000$ runs of the Q-learning algorithms are simulated to get the last $1000$ Q-values of each run and use them in a similar clustering task. More information about the clustering tasks performed is available in Appendix \ref{Appendix: detection}. Once the clusters are identified, their size is measured by counting how many of the runs ended up in each cluster. The outcome of the measurements is presented in Fig. \ref{fig: bassin_measured}.
\medbreak
The results indicate that, for a given $g$, spontaneous coupling appears for moderate values of $\varepsilon_A$ and $\varepsilon_B$. In line with Banchio and Mantegazza (2023) 's results, this region enlarges when $g$ increases, covering $0 \%$ of the square $(\varepsilon_A, \varepsilon_B) \in [0,1]$ in the case $g=1$ up to $80 \%$ of it when $g=2$. For values of $\varepsilon_A$ and $\varepsilon_B$ that are too high, spontaneous coupling doesn't appear at all, and the algorithms do not collude. No spontaneous coupling is detected at all for values of $g$ below $1.4$, which seems to contradict the theoretical result of Banchio and Mantegazza (2023). This might be due to the detection method, but in any case, the simulations indicate that spontaneous coupling is reached with negligible probability when $g$ is too low. Note also that within the existence region, the effect of $(\varepsilon_A,\varepsilon_B)$ is relatively small. Finally, one notices that asymmetry of the exploration levels is detrimental to spontaneous coupling as the probability to reach it is typically lower when $\varepsilon_A$ and $\varepsilon_B$ are very different. In such cases, for example if $\varepsilon_A>>\varepsilon_B$, under spontaneous coupling, $A$ often defects in the $\omega_{CC}$ periods causing Q-values of $B$ for action $C$ not to increase enough, which causes a switch to $\omega_{DD}$.
\medbreak
\inlinetitle{Effect on the time spent in each region} \label{sec: markov}
The previous section's objective was to measure the likelihood of spontaneous coupling. The goal of this one is to measure the time spent in each of the four regions $\{\omega_{X,Y}, X,Y \in \{C,D\}\}$ when spontaneous coupling is reached. Attention is restricted to the case $g=1.7$. For each couple $(\varepsilon_A,\varepsilon_B)$, the process is simulated $k=100$ times for $T=10^5$ periods each and the Q-values for the last $1000$ periods are saved. The frequency with which a transition from a region to another happens is measured. I focus on three quantities. First, the probability with which the algorithms remain in $\omega_{CC}$ (i.e. the probability of a transition from $\omega_{CC}$ to $\omega_{CC}$ transition conditional on being in $\omega_{CC}$). This measures the length of the $\omega_{CC}$ phases. Second, the probability of a transition from $\omega_{CD}$ to $\omega_{CD}$, measuring the length of asymmetric phases in which $A$ holds $C$ as a preferred action and $B$ holds $D$ as a preferred action. Finally, the probability of a transition from $\omega_{CC}$ to $\omega_{CD}$ conditional on a transition from $\omega_{CC}$ to $\omega_{CD}\cup \omega_{DC}$ to happen. This quantity measure how likely it is that, when algorithms enter an asymmetric region, this asymmetric region be $\omega_{CD}$.\footnote{Transitions from $\omega_{DD}$ to asymmetric regions also happen, however the results of the simulations indicate that they are not affected by exploration levels.} These measures are represented in Fig. \ref{fig: transition}.
Results indicate that the length of cooperative periods (first panel of Fig. \ref{fig: transition}) is monotonic with respect to each $\varepsilon$: an increase in one's own exploration parameter is always harmful to cooperation. When $\varepsilon$ is higher, the Q-value for $D$ is updated more often in the $\omega_{CC}$ phases, causing it to increase faster and eventually to get above that of $C$ quicker. The second and third panels of Fig. \ref{fig: transition} reveal two distinct regimes. When  $\varepsilon_A>\varepsilon_B$, the time spent in $\omega_{CD}$ is short and unaffected by the exploration policies (second panel). When a transition to an asymmetric zone happens, it is rarely to $\omega_{CD}$ (third panel). To the contrary, when $\varepsilon_A<\varepsilon_B$, the time spent in $\omega_{CD}$ is high, increasing with $\varepsilon_B$ and decreasing with $\varepsilon_A$. In this regime, transitions to asymmetric zones mostly happen to $\omega_{CD}$. When a player has a higher exploration policy than their opponent, in an $\omega_{CC}$ phase, his algorithm is quicker in realizing that $D$ dominates $C$. Before his opponent's algorithm realizes this fact as well, his own algorithm plays $D$ most of the time while the other's plays $C$ most of the time. Increasing one's own exploration policy allows one to enjoy this favorably asymmetric situation for \textit{longer} (panel 2) and \textit{more often} (panel 3). The player with lowest exploration policy can mitigate this by increasing his own level of exploration. This comes at the (social) cost of decreasing the length of cooperative phases. This explanation is confirmed by the average fraction of time spent in symmetric zones (Figure \ref{fig: symmetric}).
\subsubsection{Equilibria of the game}
The results of the previous section highlighted several effects of players' exploration policies on their payoffs and on the behavior of the algorithms. By increasing one's exploration level, a player (i) enjoys a direct positive payoff effect in the $\omega_{CC}$ phases and (ii) benefits from favorable asymmetric phases in which his own algorithm exploit the other's algorithm's cooperation. This however comes at the cost of (i) decreasing the length and frequencies of phases of mutual cooperation, and even preventing them when their level of exploration is too high and of (ii) a direct negative payoff effect in the $\omega_{DD}$ phases. This section investigates how these facts impact equilibrium levels of exploration. In order to do so, pure strategy Nash equilibria of the restricted game are computationally identified and compared with Pareto optimal profiles, defined as profiles $(\varepsilon_A,\varepsilon_B)$ that maximize the sum of the two players' payoffs.\medbreak
\inlinetitle{Methodology}
Nash equilibria are obtained by generating payoff matrices for each value of $g\in [1.5,2)$. I restrict attention to this interval and do not consider lower values of $g$ since no spontaneous coupling was detected for these values. For a given $g$, payoffs for each couple $(\varepsilon_A,\varepsilon_B)$ are generated and a payoff matrix is computed. This allows one to computationally find best response functions that are then used to identify equilibria. The generation of the payoff matrix being noisy in essence, a method to take it into account is used. $M=1000$ perturbed payoff matrices are generated by adding a small noise to the original payoff matrices and equilibria for each of the matrices are computed. A detailed explanation of the method employed is provided in Appendix \ref{Appendix: method}.
\begin{figure}[ht]
\centering
    \includegraphics[width=0.8\textwidth]{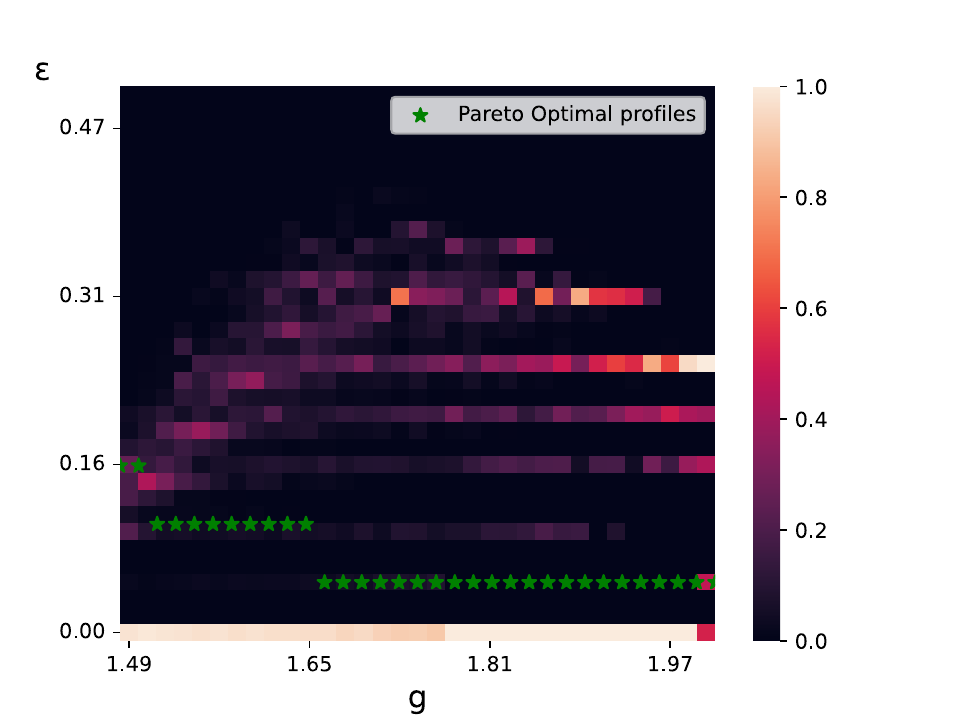}
    \caption{\textbf{Nash equilibria frequency heatmap}}
    \label{fig: eq}
\end{figure}
\medbreak
\inlinetitle{Results}
The results are shown in Figure \ref{fig: eq}. It only represents symmetric equilibria, as no asymmetric equilibria were identified. A cell corresponds to (i) a value of the parameter $g$ (ii) an interval $(\underline{\varepsilon}, \bar{\varepsilon})$. The color of the cell gives the frequency with which a symmetric Nash equilibrium of the game $G(G_0(g))$ was identified in the corresponding interval. A cell with a bright color thus indicates that a Nash equilibrium lies there with high probability. The green stars signal the Pareto profile of $G(g)$. The simulations exhibit a unique Pareto profile for each $g$, and they are all symmetric. \medbreak The first observation is that $(0,0)$ is a Nash equilibrium with high probability, coherently with Corollary \ref{cor: greedy greedy € NE}. Second, there is flagrant \textit{over-exploration} in equilibrium. This is a direct consequence of how exploration levels can be strategically increased, as explained in the previous section. Since exploration is detrimental to cooperation, collusion is thus hampered in equilibrium. However, equilibrium exploration levels cannot reach $\varepsilon=1$. Indeed, setting an exploration level too high causes the algorithms to never engage in cooperative phases. In case of the absence of cooperative phases, increasing one's level of exploration is detrimental as it has a direct negative payoff effect. Once again, Q-learning allows a form of reactive commitment. It leaves scope for mutual cooperation with phases of cooperation, but an attempt to raise one's exploration level too high to exploit these phases of cooperation is sanctioned by the cooperation disappearing. Finally, a pattern emerges when $g$ varies. First, equilibrium exploration levels increase, then they decrease as $g$ increases. The combination of two effects can explain this pattern. Raising one's exploration level is costly if it prevents phases of mutual cooperation. When $g$ is small, the exploration level cannot be increased much without cooperative phases to disappear (cf Fig. \ref{fig: bassin_measured}). As $g$ increases, players have more latitude in choosing high exploration levels and are thus incentivized to do so. However, since $u(D,C)-u(C,C)=2-g$, the incentive to benefit from the direct payoff effect in cooperative phases and from the favorably asymmetric phases decrease. On the other hand, $u(C,C)-u(D,D)=2(g-1)$ increases and phases of mutual defection become increasingly costly. When the latter effect dominates the first one, agents are incentivized to adopt lower exploration levels in equilibrium. By contrast, the socially optimal level of exploration decreases with $g$. To increase joint output, there is a trade-off in terms of direct payoff effect. In phases of mutual defection, increasing the level of exploration has a positive effect as it increases the probability that the algorithms play $C$. To the contrary, it has a negative effect in the phases of mutual cooperation. When $g$ is low, the algorithms rarely engage in spontaneous coupling. As it increases, it becomes more likely, and thus it is socially optimal to adopt lower exploration levels. Note also how Pareto optima get steady for high values of $g$, yet above $0$. This can be related to minimal exploration being required in order to maximize the probability of reaching spontaneous coupling as shown in Section \ref{sec: bassin}.
\subsubsection{Extension to $N$ players}
We now consider an extension of the parameterized prisoner's dilemma to $N\ge 2$ players. In this game, the payoffs are:
\begin{equation}
\begin{cases}
    u(C, a_{-1})=2g \frac{|\{i \in \{1...N\} \text{ s.t. } a_i=C\}|-1}{N} + g \frac{|\{i \in \{1...N\} \text{ s.t. } a_i=D\}|-1}{N} \\ u(D, a_{-1})= (2+g) \frac{|\{i \in \{1...N\} \text{ s.t. } a_i=C\}|-1}{N} + 2 \frac{|\{i \in \{1...N\} \text{ s.t. } a_i=D\}|-1}{N} 
\end{cases}
\end{equation}
This game is simply an extension of the parameterized prisoner's dilemma in which each player obtains the average payoff he would obtain when playing against each of the other players. Thus defection remains a strictly dominant strategy. We simulate the Q-learning algorithms over $T=3\times 10^5$ time periods and average the results over $M=100$ runs to computationally obtain the Nash equilibria for different values of $N$ when $\varepsilon$ can be chosen among $20$ equally spaced values in $[0,1]$. To maintain a reasonable computational cost, we restrict attention to symmetric equilibria, and when a pure-strategy Nash equilibrium fails to exist, we look for $\eta$-equilibria and compute them for the smallest value of $\eta$ such that at least one is found. We take $g=1.8$ in order to guarantee spontaneous coupling when $N=2$ with a high probability.
\begin{figure}
    \centering
    \includegraphics[width=0.9\linewidth]{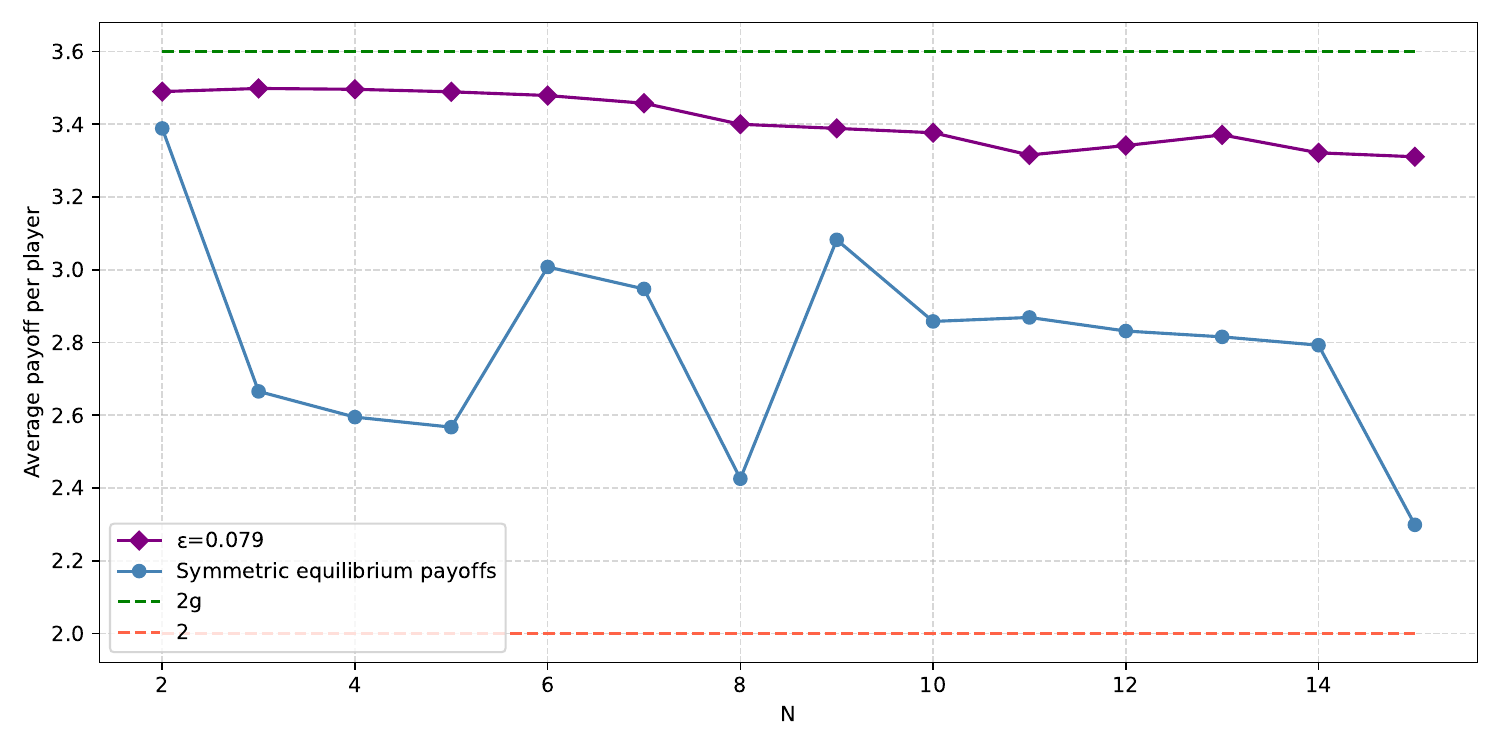}
    \caption{Equilibrium payoffs for different values of $N$}
    \label{fig: N varies}
\end{figure}
Figure \ref{fig: N varies} shows the results of our numerical experiments. The blue curve represents the average payoff per player in the symmetric equilibria. The purple curve represents the same quantity, but when all players choose the exploration level $\varepsilon=0.079$, which was the average equilibrium value of $\varepsilon$ for $N=2$. This quantity is decreasing: with a constant epsilon, it is harder to sustain cooperation, as already pointed out by Calvano et al. (2020) and as already well-known in the literature on tacit collusion (Tirole 1988, Compte et al. 2002, Ivaldi et al. 2003 \cite{tirole1988theory, compte2002capacity, ivaldi2003economics}). However it is much higher than the average payoff at the symmetric equilibrium. This reveals that increasing the number of players has an effect on their average equilibrium payoff through the equilibrium choice of the exploration level. Our interpretation of this observation is the following. As we have shown previously, a unilateral increase of one's exploration can lead to an increase in payoff but is detrimental to the overall cooperation level. When the number of players increase, the individual effect of each of them on the overall cooperation level becomes smaller. As a consequence, the individual cost of using one's own exploration level to raise one's own payoff to the detriment of overall cooperation becomes smaller, inducing a lower overall payoff in equilibrium as $N$ increases.
\section{Conclusion} \label{sec: conclusion}
In this paper, I have defined and studied a \textit{two-stage game} in which strategically chosen algorithms repeatedly play a game from a class in which actions are ranked according to their degree of cooperativeness. It includes the prisoner's dilemma, first and second-price auctions as well as Bertrand competition. The applications of such work range from online retail pricing to automated auction bidding, and closely relate to the concerns about the harmful consequences of algorithmic collusion. I proved that in such a game, any Nash equilibrium is collusive. This means that moving the margin of competition to the designing of the algorithms does not solve the problem of tacit collusion by reinforcement learning algorithms. This works points out that in order to hamper algorithmic collusion, the technology itself should be regulated. Q-learning algorithms enforce a form of \textit{reactive commitment}, which allows them to leave scope for collusive behavior without being exploited by their opponent, like a naive cooperator could be. They do not need any sort of bias towards cooperative actions to act this way. Throughout the article, I have assumed that initial conditions are drawn randomly from a well chosen interval, which we can interpret as the result of a previous period of extensive exploration. Its conclusions remain valid when players can strategically choose their initial conditions, as shown in Appendix \ref{Appendix: different intialization}. In this case, a form of bias towards cooperative actions can be introduced, which would make collusion easier. The results can also be extended when other learning parameters are strategically chosen ($\alpha$, $\gamma$). \medbreak Q-learning's ability to reactively commit has been made clearest for the special case of the greedy exploration policy. This specific policy plays a major role in proving that the Nash equilibria are collusive. It indeed allows the players to secure a collusive payoff while leaving the door open for another greedy Q-learning to be cooperative. The numerical experiments presented in section \ref{sec: simulations} show that this ability is not exclusive to the greedy policy. Exploration levels can be used strategically to exploit the other player's algorithm in phases of cooperation and to benefit from favorably asymmetric phases. As a consequence, collusion is hampered by over-exploration in equilibrium. However, an attempt to overly exploit the other algorithm by setting a level of exploration too high is sanctioned by the disappearing of cooperative phases. Once again, Q-learning algorithms feature limited tolerance and reactive commitment to cooperation. However when the number of players increase, the cooperation gets harder to sustain as the effect of each player on the overall cooperation of the algorithms gets smaller. This work also extends previous results on the behavior of $\varepsilon$-greedy algorithms in the prisoner's dilemma by means of extensive numerical simulations. They allowed to detect and measure collusive behavior as well as to provide comparative statics.
\medbreak
These results call for further work. As I have restricted the game so that players must choose a Q-learning equilibrium, it is not known how Q-learning performs when facing other types of algorithms. An investigation of this point is of primary importance to know whether Q-learning algorithms, or more advanced Q-learning-based reinforcement learning algorithms are likely to be used in practice. This article's results allow to state that, if they are, then they should be regulated to avoid their harmful consequences. The next important step is to know whether they are.\newpage
\appendix
\addcontentsline{toc}{section}{Appendices}
\section*{Appendices}
\section{Bertrand duopoly with horizontal differentiation} \label{appendix: Bertrand}
Consider a differentiated duopoly in which the demand to firm $A$ under the profile of prices $p_A, p_B$ is
\begin{equation}
    D_A(p_A,p_B)=\frac{e^{-\frac{a-p_A}{\lambda}}}{1+e^{-\frac{a-p_B}{\lambda}} + e^{-\frac{a-p_A}{\lambda}}}
\end{equation}
We shall assume players can choose prices from the same grid of prices $p^N \le  p_1...p_K \le p^M$ where $p^N$ is the Nash equilibrium price and $p^M$ the monopoly price. Condition $1$ thus holds. The profit to player $A$ under profile $(p_A, p_B)$ writes 
\begin{equation}
\begin{split}
    u(p_A,p_B)&=\frac{e^{-\frac{a-p_A}{\lambda}}}{1+e^{-\frac{a-p_B}{\lambda}} + e^{-\frac{a-p_A}{\lambda}}} (p_A-c) \\ &= \lambda \frac{e^{-\frac{p_A}{\lambda}}}{e^{-\frac{a}{\lambda}} + e^{-\frac{-p_A}{\lambda} + e^{-\frac{p_B}{\lambda}}}}\Big(\frac{p_A}{\lambda} - \frac{c}{\lambda}\Big)
\end{split}
\end{equation}
so it will be enough to show that the assumptions hold for the symmetric game in which players select prices $p_A, p_B$ from a finite grid and $A$ receives the payoff:
\begin{equation}
    u(p_A,p_B)=\frac{e^{-p_A}}{K+e^{-p_A}+ e^{-p_B}}(p_A - c)
\end{equation}
where $K, c >0$. \medbreak 
The payoff to $A$ is clearly increasing in $p_B$, so that condition $2.b.$ holds. Further consider the profit to player $A$ under the symmetric profile of prices $(p,p)$. It writes 
\begin{equation}
    u(p,p)=\frac{e^{-p}}{K+2e^{- p}} \big(p-c \big )
\end{equation}
Differentiating with respect to $p$ yields
\begin{equation}
    \frac{\partial u(p,p)}{\partial p}=\underbrace{\frac{e^{-\mu p}}{K+2e^{-\mu p}}}_{>0}\underbrace{\Big[ 1 - K \frac{p-c}{K+2e^{-\mu p}} \Big]}_{\equiv \psi(p)}
\end{equation}
where $\psi'(p)=-K \big[ \frac{K+2e^{-p}+2(p-c)e^{-p}}{(K+2e^{-p})^2} \big]<0$. Since $\psi(c)=1$ and $\lim_{+\infty} \psi(p)=-\infty$, $\frac{\partial u(p,p)}{\partial p}$ is positive then negative, so that the profit under symmetric profiles is single-peaked with maximum at $p^M$. If prices are set below $p^M$, then condition $2. a.$ is verified. \medbreak The derivative of $\pi$ with respect to $p_A$ gives:
\begin{equation}
    \frac{\partial u}{\partial p_A}=\frac{e^{-p_A}}{K+e^{-p_A}+e^{-p_B}}\Big[ 1 - \frac{(p_A-c)(K+e^{-p_B})}{K+e^{-p_A}+e^{-p_B}} \Big]
\end{equation}
so that $ \frac{\partial u}{\partial p_A}=0 \iff  \frac{(p_A-c)(K+e^{-p_B})}{K+e^{-p_A}+e^{-p_B}}=1$. The left part of the last equivalence is increasing in $p_A$, is negative when $p_A=0$ and diverges to $+\infty$ when $p_A$ goes to $+ \infty$. Hence there is exactly one solution, which we denote $p_A^*(p_B)$. Since $\frac{(p_A-c)(K+e^{-p_B})}{K+e^{-p_A}+e^{-p_B}}$ is decreasing with $p_B$, then $p_A^*$ is an increasing function. Now, observe that $\frac{\partial u}{\partial p_A}(p_A,p_B){\Big|_{p_A=c}}>0$ and $\lim_{p_A \rightarrow +\infty}\frac{\partial u}{\partial p_A}(p_A,p_B)=0$ so that $\frac{\partial u}{\partial p_A}(p_A,p_B)$ is positive, then negative. Finally, we evaluate $\frac{\partial u}{\partial p_A}(p_A,p_B)$ at $p_A=p_B=p$:
\begin{equation}
    \frac{\partial u}{\partial p_A}(p_A,p_B)\Bigg|_{p_A=p_B=p}= \frac{e^{-p}}{K+2e^{-p}}\Big[1- \underbrace{\frac{(p-c)(K+e^{-p})}{K+2e^{-p}}}_{\equiv\varphi(p)} \Big]
\end{equation}
where $\varphi$ is decreasing. Thus $ \frac{\partial u}{\partial p_A}(p_A,p_B)\Bigg|_{p_A=p_B=p}$ is decreasing, positive for $p<p^N$ and negative for $p>p^N$. Thus for $p_B>p^N$, $\frac{\partial u(p_A,p_B)}{\partial p_A}\Bigg|_{p_A=p_B}<0$. As a consequence, for all $p_A >p_B$, $\frac{\partial u(p_A,p_B)}{\partial p_A}<0$, hence no profitable deviation to a price higher than $p_B$ exists whenever $p_B>p^N$. This proves that condition $3)$ holds.
\section{Q-learning} \label{Appendix: q-learning}
Q-learning is a reinforcement learning principle designed to find optimal solutions to optimization problems in a Markov environment. Formally, denote $S$ a (finite) set of states, $A$ a set of actions and $\pi(s,a)$ the (possibly stochastic) reward obtained in state $s$ after taking action $a$. In each period, an action is taken, a reward is realized and the process moves to the next state with a probability $F(s_{t+1}|s_t, a_t) $. The objective is to learn the best policy, i.e. the one that maximizes $\mathbb{E}\Big{[}\sum_{t=0}^{+\infty}\gamma ^t \pi_t\Big{]}$ with $\gamma \in (0,1)$ a discount rate. Q-learning is an iterative method that allows one to learn the best policy function without information or hypothesis about the transition function $F$. Classically, the Bellman value function in such a problem writes as follows:
\begin{equation}
    \forall s\in S, V(s)=\mathbb{E}(\pi(s,a))+ \gamma \mathbb{E}(V(s')).
\end{equation}
The Q-matrix assigns a value to each state-action pair, and is defined as
\begin{equation}
   \forall (s,a)\in A\times S, Q(s,a)=\mathbb{E}[\pi | s,a] + \gamma \mathbb{E}[\max_{\{a' \in A\}}Q(s',a') | s, a],
\end{equation}
and linked to the Bellman value function as follows
\begin{equation}
    V(s)=\max_{a\in \mathcal{A}}Q(s,a).
\end{equation}
An agent who knows the Q-matrix exactly knows what action to take in each state, and thus knows the optimal policy. Q-learning estimates this matrix by an iterative procedure. The agent begins with an arbitrary $Q_0$ and updates any cell of the matrix she visits as follows:
\begin{equation} \label{eq: update rule}
    Q_{t+1}(s,a)=(1-\alpha) Q_t(s,a) + \alpha \Big{[} \pi_t + \gamma \max_{a'\in A} Q_t(s',a) \Big{]},
\end{equation}
where $\alpha$ is referred to as the \textit{learning rate} and controls how rapidly Q-values change when a reward is obtained. Note that when $\alpha=0$ the Q-values stay constant, so that the agent does not learn, and when $\alpha=1$ the Q-values immediately  change to the actualized reward. \medbreak
Convergence to the optimal Markov policy is guaranteed under conditions on the learning rates and on the exploration policy (Singh et al. 2000 \cite{singh2000convergence}):
\begin{proposition01}
Given a GLIE policy (Greedy in the Limit with Infinite Exploration), \textit{i.e.} such that
\begin{itemize}
    \item The exploration policy converges to the greedy one as $t$ goes to $\infty$
    \item Every action-state pair is visited infinitely often,
\end{itemize}
if the sequence of learning rates satisfy
\begin{enumerate}
    \item $\sum_{t\in \mathbb{N}}\alpha_t=+\infty$
    \item $\sum_{t\in \mathbb{N}}\alpha_t^2<+\infty$,
\end{enumerate}
then $(Q_t)_{t\in \mathbb{N}}$ converges to the Q-matrix with probability $1$.
\end{proposition01}
There are three important differences between the setting studied in this article and classical Q-learning. First, I consider \textit{stateless} Q-learning for which the update rule becomes:
\begin{equation}
\left\{
    \begin{array}{ll}
    Q_{t+1}(a)=(1-\alpha) Q_t(a) + \alpha \Big{[} \pi_t + \gamma \max_{a'\in A} Q_t(a) \Big{]} \text{ if $a$ is played at $t$}
    \\
    Q_{t+1}(a)=Q_{t}(a) \text{ otherwise}.
    \end{array}
\right.
\end{equation}
Second, learning rates are assumed to be constant and the exploration policy to be stationary. This naturally prevents any of the policy functions considered to be GLIE. Third, and most importantly, the environment faced by a Q-learning algorithm is not markovian. Q-learning algorithms' actions are affected by the \textit{whole history of play} as they take their decisions based on evolving Q-values. This fact prevents any theoretical guarantee of convergence. The algorithms are facing a \textit{moving target problem}: one adapts to the other, that in returns adapts to the first one. This well-known problem in the multi-agent reinforcement learning literature has motivated a whole strand of research concerned with the design of reinforcement learning algorithms converging to desirable outcomes when interacting together. Busoniu et al. (2008) \cite{busoniu2008comprehensive} provide a survey of this literature.
\section{Different initializations} \label{Appendix: different intialization}
\subsection{Initialization as an average}
In this section we shall assume that instead of drawing the initial conditions according to some joint distribution, the players initialize their algorithms' Q-values with an average of the actions' payoffs. Formally we shall assume:
\begin{equation}
    \forall a \in \mathcal{A}, \forall i \in \{A,B\}, Q^i_0(a)=\frac{1}{1-\gamma} \frac{1}{|\mathcal{A}|}\sum_{a' \in \mathcal{A}} u(a,a')
\end{equation}
This initialization has two possible interpretations. Either the players know the payoff matrix and initialize the Q-values of their algorithms to reflect how each action performs on average against the others. Or the players know nothing about the payoff matrix, and the algorithms start with a long period of exploration (like in Lambin (2024) \cite{lambin2024less}). We shall denote $\hat{u}(a) = \frac{1}{|\mathcal{A}|}\sum_{a' \in \mathcal{A}} u(a,a')$. \medbreak
We shall take the following assumption
\begin{assumption} \label{assumption: admissible}
    There exists $a \ne a_1$ such that $\hat{u}(a)>u(a_1,a_1)$.
\end{assumption}
That is, there is at least one action that on average gives strictly higher-than-Nash payoffs against the others. We shall show that under Assumption \ref{assumption: admissible}, Theorem \ref{Thm1} remains true. First, observe that Proposition \ref{prop: asymmetric exploration} remains true: if one of the algorithms is greedy while the other one is not, our result on the limiting behavior of the algorithms remains unchanged. Indeed, the argument of the proof of Proposition \ref{prop: asymmetric exploration} only relies on the initialization of the Q-value of $a \in \mathcal{A}$ in $I_a$, which clearly is the case when it is initialized at $\hat{u}(a)$. We now prove that Proposition \ref{prop: 00} remains valid under Assumption \ref{assumption: admissible} in order to conclude that Theorem \ref{Thm1} remains true.
\begin{proposition}
    Under Assumption \ref{assumption: admissible}, $\Pi(0,0)>u(a_1,a_1)$
\end{proposition}
\begin{proof}
    By assumption, the Q-values of the two players' algorithms are initially symmetric. Since the greedy algorithm only selects the action with the highest Q-value, both algorithms always choose the same action so that for all $a, t, Q_t^A(a)=Q_t^B(a)$, which we shall denote $Q_t(a)$. Now, observe that if the algorithms play $a_1$ at $t$: 
    \begin{equation}
    \begin{split}
        Q_{t+1}(a_1)&=(1-\alpha)Q_t(a_1)+\alpha \Big[ u(a_1,a_1) + \gamma Q_t(a_1) \Big] \\  \iff Q_{t+1}(a_1)-Q_t(a_1) &= \alpha \Big[ u(a_1,a_1) -(1-\gamma)Q_t(a_1) \Big]
    \end{split}
    \end{equation}
    where $\alpha \Big[ u(a_1,a_1) -(1-\gamma)Q_t(a_1) \Big] \le 0$ since $u(a_1,a_1) = \min_{a} u(a_1,a) $ by assumption and $Q_0(a_1) \in I_{a_1}$ so that $Q_t(a_1) \in I_{a_1}$. Thus, whenever the algorithms play action $a_1$, the corresponding Q-value decreases. \medbreak If at some $t$, the Q-value of the action played (weakly) increases, then this action will be played forever on. By the previous point, this action cannot be $a_1$, so in this case both players receive payoffs $u(a, a)>u(a_1,a_1)$. Now, let us assume that for all $t$, the Q-value of the action played at $t$ decreases. If an action $a$ is played infinitely many times, then it is immediate from the update rule that its value must converge to $\frac{u(a,a)}{1-\gamma}$. Since by assumption, for all $k$, $u(a_k,a_k)<u(a_{k+1}, a_{k+1})$, it is clear that only one action can be played infinitely many times. Now observe that for any $a \in \mathcal{A}$ such that $\hat{u}(a)> u(a_1,a_1)$, the Q-value of $a$ is monotonic. Indeed, upon updating $a$:
    \begin{equation}
    \begin{split}
        Q_{t+1}(a)&=(1-\alpha) Q_t(a) + \alpha \Big[ u(a,a) + \gamma Q_t(a) \Big] \\ Q_{t+1}(a)-Q_t(a) &= \alpha \Big[ u(a,a) -(1-\gamma)Q_t(a) \Big]
    \end{split}
    \end{equation}
    so that as long as $Q_t(a)$ is below $\frac{u(a,a)}{1-\gamma}$, it is increasing, and as long as $Q_t(a)$ is above $\frac{u(a,a)}{1-\gamma}$ it is decreasing. Thus, for all $t$, $Q_t(a) \in \big[ \min\big(\frac{u(a,a)}{1-\gamma}, \frac{\hat{u}(a)}{1-\gamma}\big), \max\big(\frac{u(a,a)}{1-\gamma}, \frac{\hat{u}(a)}{1-\gamma}\big)  \big]$. By assumption \ref{assumption: admissible}, there exists at least one $a \in \mathcal{A}$ such that $\hat{u}(a)>\frac{u(a_1,a_1)}{1-\gamma}$. Thus it cannot be the case that $\lim_{+\infty} Q_t(a_1) = \frac{u(a_1,a_1)}{1-\gamma}$, as it would imply that $Q_t(a_1)<Q_t(a)$ for $t$ large enough, implying that $a_1$ is no longer updated. As a consequence, the algorithms end up playing an action $a \ne a_1$ forever on, which gives them a payoff above $u(a_1,a_1)$.
\end{proof}
Note that we can actually precisely characterize which of the actions the algorithms will end up selecting: it is the action with highest mutual payoff among all the actions that satisfy $\hat{u}(a)>u(a_1,a_1)$. From this point, all the subsequent results follow. We give conditions under which the results hold for the three examples given in Sec. \ref{sec: One-shot game}.
\subsubsection*{Prisoner's dilemma}
In the parameterized prisoner's dilemma as in Sec. \ref{sec: simulations}, the condition $\hat{u}(C)>u(D,D)$ simply writes:
\begin{equation}
    \frac{1}{4} \Big[ 2g + g \Big] > 2 \iff g > \frac{4}{3}
\end{equation}
Thus equilibria are guaranteed to be collusive as long as $g$ is large enough, that is if the gain to cooperation is large enough.
\subsubsection*{Discretized first-price auction}
In the discretized first-price auction, it is enough that bidding the second highest of the grid gives on average more than bidding the highest on the grid gives against itself. That is 
\begin{equation}
    \frac{K-1}{K}\Big[ v-b(K-1) \Big] > \frac{v-bK}{2} \iff (v-bK)(\frac{1}{2}-\frac{1}{K})>-\frac{K-1}{K}b
\end{equation}
which is true as long as $K\ge 2$. 
\subsubsection*{Discretized Bertrand duopoly}
In the discretized Bertrand duopoly, profit functions are given by:
\begin{equation}
    u(p_A, p_B)=\frac
    {e^{-\mu p_A}}{1+e^{-\mu p_A}+e^{- \mu p_B}}p_A
\end{equation}
where prices are selected on a finite grid $A=\{p_1...p_K\}$ above the Nash equilibrium price and below the monopoly price. Assumption \ref{assumption: admissible} holds if and only if for some $p \ne p_1$:
\begin{equation}
    u(p_1,p_1) < \frac{1}{K}\sum_{k=1}^K u(p,p_k) 
\end{equation}
This is true for $K$ large enough, indeed:
\begin{equation}
   \frac{1}{K}\sum_{k=1}^K u(p,p_k) > \frac{1}{K}u(p,p_1) + \frac{K-1}{K}u(p,p_2) 
\end{equation}
where clearly $\frac{1}{K}u(p_2,p_1) + \frac{K-1}{K}u(p_2,p_2)>u(p_1,p_1)$ for $K$ large enough.
\subsection{Initial conditions chosen by agents}
If initial conditions are chosen by agents along with their exploration rates, then the results hold, subject to an adjustment. In the spirit of Compte (2023) \cite{compte}, when choosing a greedy exploration policy, $A$ and $B$ can select initial conditions favoring cooperation by setting $Q_0(a_n) = \max_{a \in \mathcal{A}} Q_0(a)>\frac{\min_a u(a_n,a)}{1-\gamma}$. 
\begin{proposition}
    In the game in which $A$ and $B$ can set their initial conditions, all equilibria give both players payoffs weakly higher than $u(a_1,a_1)$. The only Nash equilibrium in which at least a player obtains payoff exactly equal to $u(a_1,a_1)$ is such that in all periods, both algorithms only play $a_1$.
\end{proposition}
\begin{proof}
    First, observe that players can replicate any constant strategy by choosing a greedy algorithm and setting all initial conditions below $\frac{\min_{a,a'} u(a,a')}{1-\gamma}$. Then, regardless of what the other player's algorithm plays, the Q-value of the updated action increases, leading to a repeated choice next period. As a consequence, it is clear that all equilibria must give payoffs weakly higher than $u(a_1,a_1)$ for both players. \medbreak Now take an equilibrium such that both players derive payoffs $u(a_1, a_1)$. If one of the players is not greedy, then the other one can select a greedy strategy replicating the constant action $a_1$, which will give him payoff strictly higher than $u(a_1,a_1)$. As a consequence, both players are necessarily greedy. Let's assume at least one of the players, say $A$, selects initial conditions such that their algorithm does not replicate the constant action $a_1$. Then distinguish two cases. If $Q_0(a_1)<\frac{u(a_1,a_1)}{1-\gamma}$, it must be the case that for some $a \ne a_1$, $Q_0(a_1)<Q_0(a)$, so that $a$ is played forever on by this algorithm. But then by matching his initial conditions, the opponent can obtain payoff $u(a,a)>u(a_1,a_1)$, which contradicts the equilibrium condition. Assume now that $Q_0(a_1) > \frac{u(a_1,a_1)}{1-\gamma}$. Since the algorithm does not replicate the constant action $a_1$, it must be the case that for some $a \ne a_1$, $Q_0(a)>\frac{u(a_1,a_1)}{1-\gamma}$: otherwise, regardless of what the other algorithm does, $a_1$ is played in each period. Assume $B$ chooses initial conditions matching those of $A$. Then whenever both algorithms play $a_1$, the associated Q-values decrease. If they do so infinitely many times, the Q-values for $a_1$ converge to $\frac{u(a_1,a_1)}{1-\gamma}$. But since there is an action $a \ne a_1$ such that $Q_0(a) > \frac{u(a_1,a_1)}{1-\gamma}$, it must be the case that, at some point, $a$ is chosen by both algorithms. If $Q_0(a) < \frac{u(a,a)}{1-\gamma}$, then the Q-value for $a$ increases whenever it is played by both algorithms, and the algorithms settle for the profile $(a,a)$ forever on. If $Q_0(a) \ge \frac{u(a,a)}{1-\gamma}$, then it remains so forever. But then it cannot be the case that $(a_1,a_1)$ is played infinitely many times, as the Q-value for $a_1$ will eventually be lower than $\frac{u(a,a)}{1-\gamma}$. As a consequence, the algorithms must eventually synchronize on the profile $(a,a)$, which will give both players a payoff strictly higher than $u(a_1,a_1)$. This contradicts the fact that both players obtain $u(a_1,a_1)$ in equilibrium and allows us to conclude the result.
\end{proof}
When the players can choose their initial conditions, then they can choose strategies that exactly replicate a constant action. Since by definition $(a_1,a_1)$ is a strict Nash equilibrium, any profile of strategies that replicate constant action $a_1$ for both algorithms is a Nash equilibrium. However, this is the only case in which one of the players make exactly $u(a_1,a_1)$ in equilibrium: in all other cases, payoffs in equilibrium are strictly collusive. Any equilibrium that is not strictly collusive does not make any use of the Q-learning algorithm and should be considered pathological.

\section{Omitted proofs}
\textbf{Proof of Proposition 1}
Focus on one of the two algorithms. First, it is clear from the update rule that if for some $t$, for all $a$, $Q_t(a) \in I_a$ then for all $t'\ge t, \forall a, Q_{t'}(a)\in I_a$.
\medbreak 
Now, assume at $t$, $\max_a Q_t(a)>\frac{\bar{u}}{1-\gamma}$. Denote $a^*_t$ the action with maximal Q-value at $t$. If it is updated while the other algorithm plays action $a'$, then the update rule gives:
\begin{equation}
    Q_{t+1}(a^*_t) - Q_t(a^*_t) = \alpha \Big{(}u(a^*_t,a') - (1-\gamma)\max Q_t(a^*_t)\Big{)} \le 0.
\end{equation}
If an action $\tilde{a}$ such that $Q_t(\tilde{a})\ne \max Q_t(a)$ is updated at $t$, then:
\begin{equation}
    Q_{t+1}(\tilde{a})\le \alpha Q_t(a^*_t) + (1-\alpha)\Big{(}(1-\gamma)Q_t(a^*_t) + \gamma Q_t(a^*_t)\Big{)}=Q_t(a^*_t).
\end{equation}
Thus, as long as $\max_{a''} Q_t(a'')$ is greater than $\frac{\max_{a',a''}u(a',a'')}{1-\gamma}$, $\max_{a''} Q_t(a'')$ is decreasing. Assume this is true for all subsequent $t$s. Then $\max_{a''}Q_t(a'')$ converges to $l\ge \frac{\max_{a', a''} u(a',a'')}{1-\gamma}$. This is only possible in two cases:
\begin{enumerate}
    \item The maximal Q-value is updated only finitely many times. By assumption, this happens with probability $0$.
    \item The maximal Q-value is updated infinitely many times, but when doing so, the algorithm received a payoff lower than $\bar{u}$ finitely many times. Since by assumption the opponent explores with probability bounded away from zero, this happens with probability $0$.
\end{enumerate}
Thus with probability $1$, there exists $t$ such that for all $a$, $Q_t(a)<\frac{\max_{a',a''} u(a',a'')}{1-\gamma}$. Note that for all subsequent $t$s, for all $a, Q_t(a) \le  \frac{\max_{a',a''} u(a',a'')}{1-\gamma}$. \medbreak Now assume that $Q_t(a)<\frac{\min_{a'} u(a,a')}{1-\gamma}$. If $a$ is updated at $t$ while the opponent plays $a'$ then:
\begin{equation}
\begin{split}
    Q_{t+1}(a)&=(1-\alpha) Q_t(a) + \alpha \Big[ u(a,a') + \gamma \max_{a''}Q_t(a'')  \Big] \\ &\ge (1-\alpha) Q_t(a) + \alpha \Big [ u(a,a') + \gamma Q_t(a) \Big]
\end{split}
\end{equation}
so that 
\begin{equation}
    Q_{t+1}(a) - Q_t(a) =\alpha \Big[ u(a,a') - (1-\gamma) Q_t(a) \Big] \ge 0
\end{equation}
Thus as long as $Q_t(a)<\frac{\min_{a'} u(a,a')}{1-\gamma}$, $Q_t(a)$ keeps increasing whenever it is updated. By a reasoning similar than for the previous point, we conclude that with probability $1$ there exists $t$ such that $Q_t(a) \ge \frac{\min_{a'}u(a,a')}{1-\gamma}$. It is also clear from the update rule that when this happens, then for all subsequent $t'$, $Q_{t'}(a) \ge \frac{\min_{a'}u(a',a)}{1-\gamma}$. \medbreak Now assume that for all $a$, $Q_t(a)\le \frac{\max_{a',a''}u(a',a'')}{1-\gamma}$ and $Q_t(a) \ge \frac{\min_{a'} Q_t(a,a')}{1-\gamma}$. Then, when action $a$ is updated at $t$, by the update rule:
\begin{equation}
\begin{split}
    Q_{t+1}(a)&=(1-\alpha)Q_t(a) + \alpha \Big[ u(a,a') + \gamma \max_{a''} Q_t(a'') \Big] \\ &\le (1-\alpha) Q_t(a)+ \alpha \Big[ \max_{a''}u(a,a'') + \gamma \max_{a',a''} \frac{u(a',a'')}{1-\gamma}\Big] 
\end{split}
\end{equation}
thus as long as $Q_t(a) > \max_{a''}u(a,a'') + \gamma \max_{a',a''} \frac{u(a,a'')}{1-\gamma}$, $Q_t(a)$ decreases each time it is updated. By a similar argument as previously, we conclude that with probability $1$, there exists $t$ such that $Q_t(a) \le \max_{a',a''}u(a',a'') + \gamma \max_{a',a''} \frac{u(a,a'')}{1-\gamma}$. This completes the proof. \qedsymbol
\medbreak

\textbf{Proof of Lemma 1}
When the action with maximal Q-value, denoted $a^{*}_t$ is updated at a period $t$, the update rule writes:
    \begin{equation*}
    \begin{split}
        Q_{t+1}(a_t^*)&=(1-\alpha) Q_t(a_t^*) + \alpha \Big[u(a^*_t, a) + \gamma Q_t(a_t^*)\Big] \\ & < (1-\alpha) Q_t(a_t^*) + \alpha \Big[u(a^*,a) + \gamma Q_t(a^*_t)\Big]
    \end{split}
    \end{equation*}
    Denote $\max Q_K$ the maximal Q-value after it has been updated $K$ times after period $t$. From the previous inequality and a trivial induction 
    \begin{equation*}
        \max Q_K < u_K 
    \end{equation*}
    where $(u_k)_{k \in \mathbb{N}}$ is defined recursively as $u_0=\max_{a'} Q_t(a')$ and $u_{k+1}=(1-\alpha) u_k + \alpha \Big[u(a^*,a) + \gamma u_k\Big]$. Further, $(u_k)_{k\in \mathbb{N}}$ converges to $\frac{u(a^*,a)}{1-\gamma}$ so that for all $\eta >0$, necessarily, there exists $K_\eta$ such that for all $k>K_\eta$, $\max Q_k <  \frac{u(a^*,a)}{1-\gamma} + \eta$. Thus, if at $t$ the action with maximal Q-value has been updated $K_\eta$ times, for all $a''\in A, Q_t(a'')<\frac{u(a^*,a)}{1-\gamma}+ \eta$. Now, observe that when $a^*$ is updated, the update rule writes:
    \begin{equation*}
    \begin{split}
        Q_{t+1}(a^*)&=(1-\alpha) Q_t(a^*) + \alpha \Big[ u(a^*,a) + \gamma \max_{a''}Q_t(a'') \Big] \\ & \ge (1-\alpha) Q_t(a^*) +  \alpha \Big{[} u(a^*,a) + \gamma Q_t(a^*) \Big{]}
    \end{split}
    \end{equation*}
    By a similar argument as previously, for any $\eta$, there exists $N_\eta$ such that if at $t$, $a^*$ has been updated $N_\eta$ times then:
    \begin{equation*}
        Q_{t}(a^*) > \frac{u(a^*,a)}{1-\gamma} - \eta
    \end{equation*}
    \medbreak 
    Thus, if the action with maximal Q-value has been updated $K_\eta$ times and $a^*$ has been updated $N_\eta$ times, then for all $a'' \in A$
    \begin{equation*}
        \frac{u(a^*,a)}{1-\gamma}-\eta<Q_t(a'')<\frac{u(a^*,a)}{1-\gamma}+\eta
    \end{equation*}
    now consider any $a'' \ne a^*$, and assume it is updated at some $t$. Then:
    \begin{equation*}
    \begin{split}
        Q_{t+1}(a'')&< (1-\alpha)Q_t(a'') + \alpha \Big( u(a'',a) + \gamma(\frac{u(a^*,a)}{1-\gamma} + \eta) \Big) \\&= (1-\alpha) Q_t(a'') + \alpha \Big( \frac{(1-\gamma)u(a'',a)+\gamma u(a^*,a)}{1-\gamma} + \gamma \eta \Big)
    \end{split}
    \end{equation*}
    By a similar reasoning as previously, after being updated $K$ times, the Q-value for $a''$ is bounded above by a sequence converging to $ \frac{(1-\gamma)u(a'',a)+\gamma u(a^*,a)}{1-\gamma} + \gamma \eta$. For $\eta< \frac{u(a^*, a)-\max_{a'' \ne a^*} u(a'',a) }{1+\gamma}$, then $ \frac{(1-\gamma)u(a'',a)+\gamma u(a^*,a)}{1-\gamma} + \gamma \eta < \frac{u(a^*, a)}{1-\gamma} -\eta$. Hence if enough updates of each action $a'' \ne a^*$ and of the action with maximal Q-value have taken place at $t$, then for all $a''$, and all $t'>t$, $Q_{t'}(a'')< Q_{t'}(a^*)$. Since the algorithm has a strictly positive exploration parameter, with probability $1$ there exists a $t$ such that this is the case. \qedsymbol

\textbf{Proof of Proposition 2}
In order to prove this result, we need the following lemma
\begin{lemma}
    Consider a greedy Q-learning algorithm with the initial Q-value for action $a \in A$ selected in $\Big[ \frac{u (a, a_1)}{1-\gamma}, \frac{u(a,a_K)}{1-\gamma}\Big]$. There exists a finite integer $L$ such that if the opponent plays $a_1$ for $L$ consecutive times, then for all $a \ne a_1, Q_a < \frac{u(a_1,a_1)}{1-\gamma}$. 
\end{lemma}
\begin{proof}
Under the greedy Q-learning algorithm, only the action with current maximal Q-value is updated. When the opponent plays $a_1$, the update rule yields 
\begin{equation*}
    Q_{t+1}(a)=(1-(1-\gamma)\alpha)Q_t(a)+\alpha u(a,a_1)
\end{equation*}
After the opponent has played $a_1$ while the greedy algorithm played action $a$ $n$ consecutive times, the Q-value for action $a$ verifies 
\begin{equation*}
    Q_n(a)=\big(1-(1-\gamma)\alpha\big)^n \Big(Q_0(a) - \frac{u(a,a_1)}{1-\gamma} \Big) + \frac{u(a,a_1)}{1-\gamma}
\end{equation*}
where $Q_0(a)$ denotes the Q-values for action $a$ before the sequence of $n$ $(a, a_1)$ profiles took place. Observe that $\lim_{+\infty}Q_n(a)=\frac{u(a,a_1)}{1-\gamma}$ and that
\begin{equation*}
    Q_n(a) \le \big(1-(1-\gamma)\alpha\big)^n \Big(\frac{u(a,a_K)-u(a,a_1)}{1-\gamma}\Big) + \frac{u(a,a_1)}{1-\gamma} \equiv u_n^a
\end{equation*}
Since $(a_1,a_1)$ is a strict Nash equilibrium, for all $a \ne a_1, u(a_1,a_1)>u(a,a_1)$ so that $\frac{u(a_1,a_1)}{1-\gamma}>\frac{u(a,a_1)}{1-\gamma}$. Thus, for $a \ne a_1$ there exists $n^*(a)=\max\Big(\{n / u_n^a < \frac{u(a_1,a_1)}{1-\gamma}\}\Big)$. Note that by definition $u^{a_1}_{n^*(a)}>\frac{\pi(a_1,a_1)}{1-\gamma}$, so that there exists $m^*(a)=\min \Big(\{ n / u_n^{\overline{a}} < u_{n^*(a)-1}^a\} \Big)$. But then if the opponent plays action $a_1$ for  $L=\sum_{a \in A -\{a_1\}} n^*(a) + \max_{a \in A - \{a_1\}} m^*(a)$ consecutive times, necessarily for all $a \ne a_1, Q(a)<\frac{u(a_1, a_1)}{1-\gamma}\le Q(a_1)$.
\end{proof}

Denote $a_t$ the action played at $t$ by $A$'s algorithm. Denote $E$ the event $(\exists t \text{ s.t. } \forall t'>t, a_t=a_1)$ and $E'$ the event $(\exists t \text{ s.t. } \forall t' \ge t, \forall a\ne a_1,  Q_{t'}^B(a_1)>Q_{t'}^B(a)$).
   \begin{equation*}
   \begin{split}
     \Pi(\varepsilon_A, \varepsilon_B)&= \mathbb{E}_{\varepsilon_A=0, \varepsilon_B>0} \Big[ \liminf _{T \longrightarrow +\infty} \frac{1}{T} \sum_{t=0}^T u(a_t, a_t') \Big]\\& = \mathbb{E}_{\varepsilon_A=0, \varepsilon_B>0} \Big[ \liminf _{T \longrightarrow +\infty}\frac{1}{T} \sum_{t=0}^T u(a_t, a_t') \Big | E\Big] \mathbb{P}(E) \\ &  + \mathbb{E}_{\varepsilon_A=0, \varepsilon_B>0} \Big[ \liminf _{T \longrightarrow +\infty}\frac{1}{T} \sum_{t=0}^T u(a_t, a_t') \Big | \neg E \Big] \mathbb{P}( \neg E)
    \end{split}
   \end{equation*}
   where $\mathbb{P}(E)=1$ (Lemma 2). Hence 
   \begin{equation*}
   \begin{split}
   \Pi(\varepsilon_A, \varepsilon_B)= &\mathbb{E}_{\varepsilon_A=0, \varepsilon_B>0} \Big[ \liminf _{T \longrightarrow +\infty} \frac{1}{T} \sum_{t=0}^T u(a_t, a_t') | E, E' \Big] \mathbb{P}(E' | E) + \\ &\mathbb{E}_{\varepsilon_A=0, \varepsilon_B>0} \Big[ \liminf _{T \longrightarrow +\infty} \frac{1}{T} \sum_{t=0}^T u(a_t, a_t') | E, \neg E' \Big] \mathbb{P}( \neg E' | E)
   \end{split}
   \end{equation*}
   where $\mathbb{P}(E' | E)=1$ (Lemma 1). So that
   \begin{equation*}
    \Pi(0, \varepsilon_B)= \mathbb{E}_{\varepsilon_A=0, \varepsilon_B>0} \Big[ \liminf _{T \longrightarrow +\infty}\frac{1}{T} \sum_{t=0}^T u(a_1, a_t') | E, E' \Big]
   \end{equation*}
   Conditional on the $E$ and $E'$, for $t$ large enough, the algorithm of player $B$ plays $a_1$ with probability $1- \frac{(K-1)\varepsilon_B}{K}$ and each $a\ne a_1$ with probability $\frac{\varepsilon_B}{K}$. Hence by the strong law of large numbers, with probability $1$:
   \begin{equation*}
       \liminf _{T \longrightarrow +\infty}\frac{1}{T} \sum_{t=0}^T u(a_1, a_t') = \Big(1-\frac{(K-1) \varepsilon_B}{K}\Big) u(a_1, a_1) + \frac{\varepsilon_B}{K}\sum_{a\ne a_1} u(a_1, a)
   \end{equation*}
   and thus, from assumptions 1 and 2, $\Pi(0, \varepsilon_B)=\Big(1-\frac{(K-1) \varepsilon_B}{K}\Big) u(a_1, a_1) + \frac{\varepsilon_B}{K}\sum_{a\ne a_1} u(a_1, a) > u(a_1,a_1)$. Symmetrically $\Pi(0, \varepsilon_B)=\Big(1-\frac{(K-1) \varepsilon_B}{K}\Big) u(a_1, a_1) + \frac{\varepsilon_B}{K}\sum_{a\ne a_1} u(a, a_1) < u(a_1,a_1)$.
\medbreak
\textbf{Proof of Proposition 3}
For $\varepsilon_A=\varepsilon_B=0$, the Q-values are deterministic given the initial conditions. Further, when an action is updated, it needs to be the action with the highest Q-value, so that, if action $a$ is updated at $t$:
    \begin{equation*}
        Q_{t+1}(a)=(1-\alpha)Q_{t}(a)+\alpha \Big[ u(a, a'_t) + \gamma Q_t(a) \Big]
    \end{equation*}
    where $a'_t$ denotes the action chosen by the opponent's algorithm at $t$. For a player $X \in \{A,B\}$ denote $\max_2 Q_t^X = \max_{a \ne \arg\max_a' Q_t^X(a')}Q_t^X(a)$ the second highest Q-value. The sequence $\Big(\max_2Q_t^X\Big)_{t\in \mathbb{N}}$ is decreasing. Indeed, since the greedy algorithm only updates the action with highest Q-value, $\max_2 Q_t^X \ne \max_2 Q_{t+1}^X$ only if the action with highest Q-value changes at $t$. This can only happen when the current maximal Q-value undercuts the second highest, which then becomes the highest. Further, $\Big(\max_2 Q_t^X\Big)_{t\in \mathbb{N}}$ is bounded below by $\frac{\min_{a_1,a_2} u(a_1,a_2)}{1-\gamma}$, hence it converges. Denote $l^X$ its limit and let us distinguish two cases.
    \begin{enumerate}
        \item If for a player, say $A$,$l^A<\frac{u(a_1,a_1)}{1-\gamma}$ then there exists $t$ such that for all $t'>t$ and all actions $a \ne \arg \max_{a'} Q_{t'}^X(a')$, $Q_t^X(a)<\frac{u(a_1,a_1)}{1-\gamma}$ and thus $Q_{t'}^X(a)<Q_{t'}^X(a_1)$. Thus from $t$ on player $X$'s algorithm only plays action $a_1$. Denote $a'_t$ the action played (and therefore updated) by $B$'s algorithm. The update rule gives 
        \begin{equation*}
        \begin{split}
            &Q^B_{t+1}(a_t')=(1-\alpha)Q_t^B(a_t') + \alpha \Big( u(a_t', a_1) + \gamma Q_t^B(a_t') \Big) \\ \iff & Q^B_{t+1}(a'_t) - Q^B_t(a_t') = \alpha \Big[  u(a_t',a_1) - (1-\gamma) Q_t^B(a') \Big] < \alpha \Big[  u(a_1,a_1) - (1-\gamma) Q_t^B(a') \Big] 
        \end{split}
        \end{equation*}
        so that the Q-value of updated actions is decreasing. Let $a \ne a_1$ and assume $a$ is updated infinitely many times. Then $\lim_{+\infty} Q^B_t(a)< \frac{u(a_1,a_1)}{1-\gamma}$, which yields a contradiction. As a consequence, every action that is not $a_1$ must be played finitely many times, and hence there exists a time after which $B$'s algorithm only plays $a_1$. Denote $a_t$ the action selected by $A$'s algorithm at $t$ and $a_t'$ the action selected by $B$'s algorithm at $t$. In such a realization $\liminf_{T \rightarrow +\infty} \frac{1}{T} \sum_{t=0}^{T} u(a_t, a_t')=u(a_1,a_1)$.
        \item Else, for both players $l^X \ge \frac{u(a_1,a_1)}{1-\gamma}$. Consider an action $a$ that is played infinitely many often by $A$'s algorithm. Denote $t_k$ the time at which action $a$ was updated for the $k$-th time. The Q-value of $a$ doesn't change between two updates. Thus the update rule writes:
        \begin{equation*}
            Q_{t_{k+1}}^A(a)-Q_{t_k}^A(a)=\alpha \Big(u(a, a_{t_{k}}) - (1-\gamma) Q_{t_k}^A(a)\Big)
        \end{equation*}
        so that taking the average of the previous equation over the $K$ first updates of $a$ yields:
        \begin{equation*}
            \frac{Q_{t_K}-Q_{t_0}}{K}=\alpha \Big[ \frac{1}{K}\sum_{k=0}^{K-1}u(a,a_{t_k}) - \frac{(1-\gamma)}{K} \sum_{t=0}^{K-1}Q_{t_k}(a) \Big] \le \alpha \Big[ \frac{1}{K}\sum_{k=0}^{K-1}u(a,a_{t_k}) - \frac{(1-\gamma)}{K} K \frac{u(a_1,a_1)}{1-\gamma}     \Big]
        \end{equation*}
        where the inequality comes from the fact that $\max_2Q_t^A$ is decreasing and converges to $l^A\ge \frac{u(a_1,a_1)}{1-\gamma}$ so that an action that is updated necessarily has Q-value above $\frac{u(a_1,a_1)}{1-\gamma}$. Taking $K$ to infinity in the previous inequality gives:
        \begin{equation*}
            \liminf_{K \rightarrow +\infty} \frac{Q_{t_k}-Q_{t_0}}{K} \le \alpha \liminf_{K \rightarrow +\infty} \frac{1}{K} \sum_{k=0}^{K-1}u(a,a_{t_k}) - u(a_1,a_1)
        \end{equation*}
        so that $\liminf_{K \rightarrow +\infty} \frac{1}{K} \sum_{k=0}^{K-1}u(a,a_{t_k}) \ge u(a_1,a_1)$. Thus, any action that is played infinitely often yields an average payoff that is higher than $u(a_1,a_1)$. Now denote $t^*$ the first time such that no action that is updated finitely many time is selected and observe that 
        \begin{equation*}
            \liminf_{t \rightarrow +\infty} \frac{1}{T} \sum_{t=0}^{T}u(a_t, a'_t) = \liminf_{t\rightarrow +\infty}\frac{1}{T-t^*+1}\sum_{t=t^*}^{T}u(a_t,a_t')
        \end{equation*}
        Denote $A^*$ the set of actions that are updated after $t^*$ and for $a$ in $A^*$ denote $t_k(a)$ the time at which action $a$ is updated for the $k$-th time \textit{after} $t^*$. The right hand side of the equation writes
        \begin{equation*}
        \begin{split}
        \liminf_{t\rightarrow +\infty}\frac{1}{T-t^*+1}\sum_{t=t^*}^{T}u(a_t,a_t') &= \liminf_{K \rightarrow\infty} \frac{1}{K |A^*|} \sum_{k=0}^{K} \sum_{a\in A^*}u(a,a'_{t_k}) \\&\ge \frac{1}{|A^*|}\sum_{a \in A^*}\liminf_{K \rightarrow +\infty} \frac{1}{K}\sum_{k=0}^{K}u(a,a_{t_k}) \text{ (By superadditivity of $\liminf$)} \\ & \ge u(a_1,a_1)
        \end{split}
        \end{equation*}
    \end{enumerate}
    Thus, for any initial condition, when the two algorithms are greedy the sequence of payoffs they collect has a limit inferior higher than $u(a_1,a_1)$. Note that if the algorithms begin by playing action $K$, then the associated Q-values increase, and thus they keep playing $a_K$ forever on. The average payoffs they collect in such a case is $u(a_K, a_K)>u(a_1,a_1)$. Observing that this happens with strictly positive probability completes the proof.
\section{Numerical experiments: methods} 
 \label{Appendix: method}
\subsection{Time heatmaps and payoff functions}
In order to get the time spent in each region $\omega_{X,Y}$, the Q-learning algorithms are simulated over $T=10^5$ periods for every triplet of parameters $(g,\varepsilon_A, \varepsilon_B)$ on a grid of size $64 \times 64 \times 64$. Q-values for the last $T'=1000$ periods are saved, which allows one to deduce which region algorithms are in over this period of time. This process is reproduced $k=100$ times, and results on the share of time spent in each region are averaged over all the runs. Using the same simulations allows one to get the transitions between regions as presented in Sec. \ref{sec: markov}.\medbreak Using the measured share of time spent in each $\omega_{XY}$, payoffs attributed to each triplet of parameters are obtained using the following formula:
\begin{equation}
\Pi^{A}(\varepsilon_A,\varepsilon_B)=\sum_{(X,Y)\in \{C,D\}^2}\tau_{X,Y}(\varepsilon_A,\varepsilon
_B)u^A_{X,Y}(\varepsilon_A,\varepsilon_B),
\end{equation}
where 
\begin{equation}
        \left\{
    \begin{array}{ll}
    u_{C,C}^A=(1-\frac{\varepsilon_A}{2})(1-\frac{\varepsilon_B}{2}).2g+(1-\frac{\varepsilon_A}{2})\frac{\varepsilon_B}{2}.g+\frac{\varepsilon_A}{2}(1-\frac{\varepsilon_B}{2}).(2+g)+\frac{\varepsilon_A}{2}\frac{\varepsilon_B}{2}.(2)
        \\
    u_{C,D}^A=(1-\frac{\varepsilon_A}{2})(1-\frac{\varepsilon_B}{2}).g+(1-\frac{\varepsilon_A}{2})\frac{\varepsilon_B}{2}.2g+\frac{\varepsilon_A}{2}(1-\frac{\varepsilon_B}{2}).2+\frac{\varepsilon_A}{2}\frac{\varepsilon_B}{2}.(2+g)
        \\
    u_{D,C}^A=(1-\frac{\varepsilon_A}{2})(1-\frac{\varepsilon_B}{2}).(2+g)+(1-\frac{\varepsilon_A}{2})\frac{\varepsilon_B}{2}.2+\frac{\varepsilon_A}{2}(1-\frac{\varepsilon_B}{2}).2g+\frac{\varepsilon_A}{2}\frac{\varepsilon_B}{2}.g
        \\
    u_{D,D}^A=(1-\frac{\varepsilon_A}{2})(1-\frac{\varepsilon_B}{2}).2+(1-\frac{\varepsilon_A}{2})\frac{\varepsilon_B}{2}.(2+g)+\frac{\varepsilon_A}{2}(1-\frac{\varepsilon_B}{2}).g+\frac{\varepsilon_A}{2}\frac{\varepsilon_B}{2}.2g
    \end{array}
\right.
    \end{equation}
This allows one to get, for each $g$ on the grid, a payoff matrix of size $(64\times 64)$. Using this method to generate a payoff matrix rather than getting it directly from the simulations allows one to minimize noise in the results, however both method return similar results (with, indeed, more noise when payoffs are measured directly). Note that this method allows one to interpolate: assuming the variation of $\tau_{XY}$ with respect to $\varepsilon_A$ and $\varepsilon_B$ is low enough, one can use the results obtained on the $64\times 64 \times 64$ grid to interpolate the payoff function.
\subsection{Equilibria}
\begin{figure}
    \centering
    \includegraphics[width=0.49\textwidth]{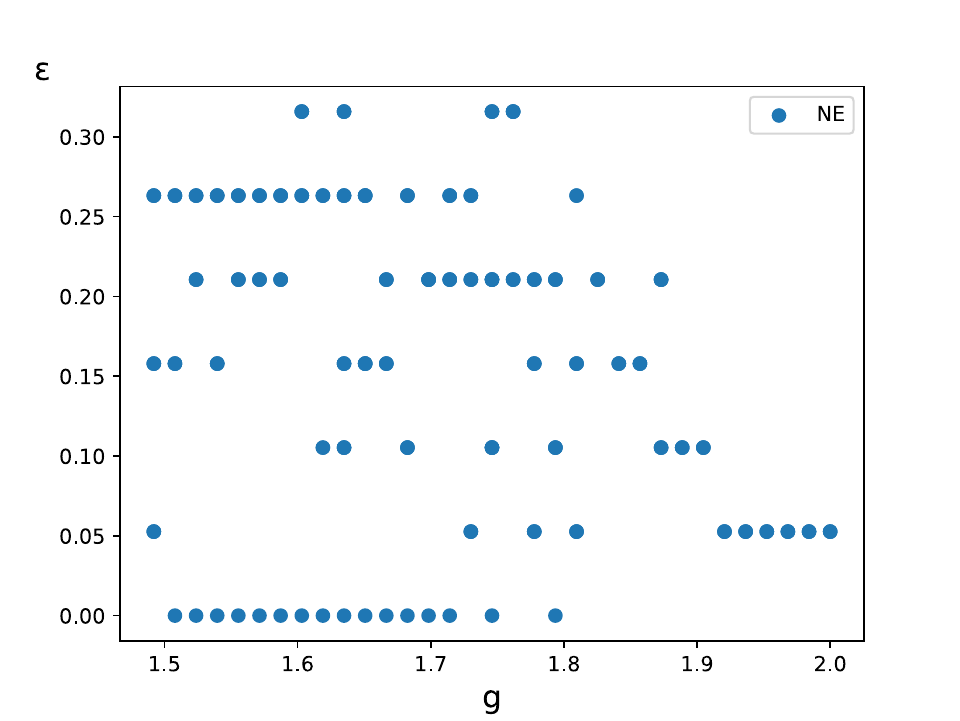}
    \includegraphics[width=0.49\textwidth]{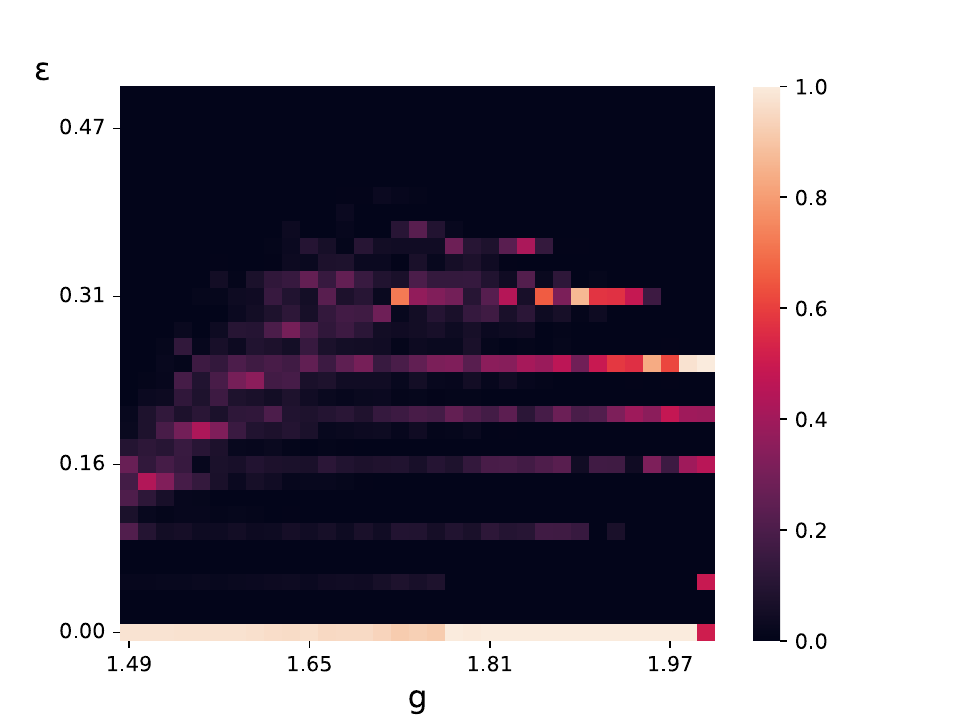}
    \caption{\textbf{Left panel:} Nash equilibria obtained from the payoff matrix \textbf{Right panel:} Nash-equilibria frequency heatmap. }
    \label{fig:comparison NE}
\end{figure}

Once payoff matrices are generated for each $g$, best response functions are computed. This requires the mere assumption that best responses are single-valued (which the simulations' results confirm). At this stage, a best response function is a (finite) list giving, for each $\varepsilon_B$, the best response $\varepsilon_A^*$ of player $A$. The list is then interpolated and best response functions are generated as a collection of linear functions (between each $\varepsilon_B$ on the grid). For a given $g$, the best response function is stored as a list of couples $(a,b)$ characterizing the best response function between two $\varepsilon_B$s on the grid. This allows one to then construct the symmetric of the curve of the best response function with respect to the $45°$ line, and to subsequently find Nash equilibria as a profile $(\varepsilon_A, \varepsilon_B)$ lying at the intersection of the best response function and its symmetric. This method allows one to get Fig. \ref{fig:comparison NE}'s left panel.\footnote{This method can of course be used since $G(g)$ is a symmetric game.}
\subsection{Denoised equilibria}
The aforementioned method generates a noisy outcome since the $\tau_{XY}$s are generated by a noisy process. In order to take this into account, $M=1000$ new observations for each $\tau_{XY}^g(\varepsilon_A,\varepsilon_B)$ are generated by randomly adding noise to the originally obtained $\tau_{XY}$s. To generate each new observation, for each triplet of parameters, a region is picked at random and the time spent inside is increased by $\eta=0.005$, then, a different region is picked at random and the time spent inside is decreased by the same amount. \footnote{Only regions that allow for the time spent inside to be increased/decreased are considered for being picked} This process generates $M$ new tables of time spent in each region for each triplet of parameters, which are used to generate the same number of payoff matrices. By using the same process as previously described on each of these payoff matrices,$M$ new lists of equilibria are obtained (one per perturbed table). To obtain Fig. \ref{fig:comparison NE}'s right panel, $(g,\varepsilon)$ are displayed on a $(64\times 64)$ grid. For $(k,m)$ on the grid, denote $g_k$ and $\varepsilon_m$ the corresponding values of $g$ and $\varepsilon$. In each cell $(k,m)$ the frequency with which there is a Nash equilibrium in the game $G(g_k)$ with perturbed payoffs lying between $\varepsilon_k$ and $\varepsilon_{k+1}$ is represented by means of a color. This method thus allows one to get where equilibria are most likely to be found and to account for the noise in the results obtained previously.
\section{Detection of spontaneous coupling: method} \label{Appendix: detection}
\begin{figure}[ht]
        \begin{minipage}[b]{0.32\linewidth}
            \centering
            \includegraphics[width=\textwidth]{2D_representation_g=1.7eA0.10526315789473684eB0.10526315789473684.pdf}
            \caption*{$g=1.7, \varepsilon_A=\varepsilon_B=0.1$}
        \end{minipage}
        \begin{minipage}[b]{0.32\linewidth}
            \centering
        \includegraphics[width=\textwidth]{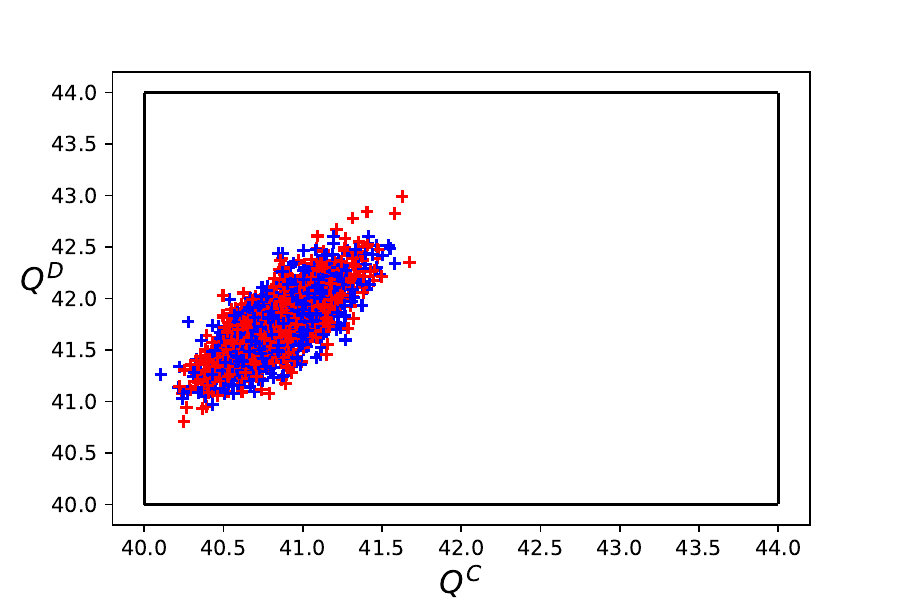}
            \caption*{$g=1.1, \varepsilon_A=\varepsilon_B=0.3$}
        \end{minipage}
        \begin{minipage}[b]{0.32\linewidth}
            \centering
        \includegraphics[width=\textwidth]{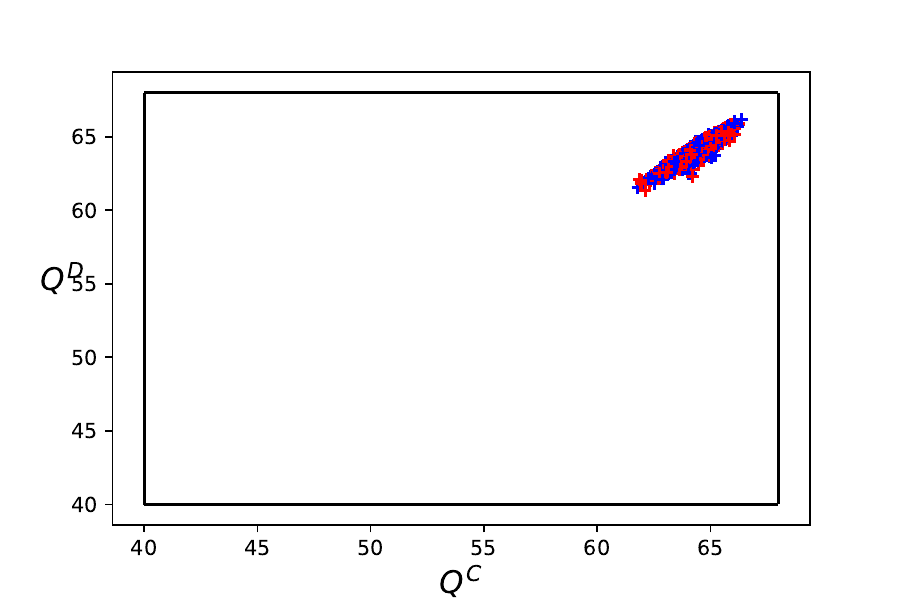}
            \caption*{$g=1.7, \varepsilon_A=\varepsilon_B=0.3$}
        \end{minipage}
    \caption{Projection on a plane of the position of the algorithms in the $4$ dimensional space}
    \label{fig: position}
    \end{figure}

In order to detect spontaneous coupling and measure its probability of appearance, the position of Q-values in the four-dimensional space $(Q^A_C, Q^A_D, Q^B_C, Q^B_D)$ after a large enough number of periods is saved. Fig. \ref{fig: position} shows the projection of the position of the algorithms on a plane for $K=1000$ different initial conditions uniformly drawn at random in the appropriate intervals: a red dot corresponds to the position of player's $B$ Q-values for one initial condition, while a blue dot corresponds to that of player $A$. The black lines correspond to values $\frac{2}{1-\gamma}$ and $\frac{2g}{1-\gamma}$, for which $Q_C$ can be above $Q_D$. On the first panel ($g=1.7, \varepsilon_A=\varepsilon_B=0.1$), two clusters clearly appear. The one on the top right part of the figure corresponds to the spontaneous coupling being reached, while the second corresponds to the algorithms remaining in $\omega_{DD}$. The simple idea behind the measure of the probability that spontaneous coupling appears relies on automatically detecting those clusters and measuring their sizes. To do so, a simple K-means algorithm with $K=2$ is used to detect the clusters and label the points accordingly.\footnote{This task is performed in the $4$-dimensional space, not in the plane, that is used for representation purposes only.}
\medbreak
It is important to note that K-means requires the number of clusters to be set in advance, and thus cannot by itself find the optimal number of clusters to perform its task. Here, this causes issues for two cases. The first case is the one in which spontaneous coupling does not appear and thus only one cluster can be found. The second panel of Fig. \ref{fig: position} shows an example where this happens. The second case is the one in which algorithms always reach spontaneous coupling in the results of the experiments. In this case too, only one cluster appears (see the third panel of Fig. \ref{fig: position} for an example). An attempt to measure directly the size of the clusters in these cases would cause the K-means algorithm to identify two clusters of equal size and thus give a wrong measure. In order to circumvent this difficulty, two additional clustering tasks using K-means algorithms with $K=2$ are performed. The first one aims at identifying the parameters which allow for spontaneous coupling. For this purpose, the time spent in $\omega_{CC}$ is used as a feature and a K-means algorithm in one dimension is run. As it is known in advance that the results should display two clusters (i.e. allowing for spontaneous coupling or not allowing for spontaneous coupling) there is no problem in fixing beforehand the number of clusters to be $2$. This way, parameters allowing for spontaneous coupling are successfully identified. For the others, we fix the measure of the probability to reach spontaneous coupling is set to $0$. The previously described clustering task is then performed. It outputs two metrics. First, the difference between the inertia metric associated to this task's outcome as well as the inertia metric associated to the trivial task with $K=1$. Second, the distance between the identified clusters' centers. These two features are used to perform yet another K-means clustering with $K=2$ on the parameters allowing for spontaneous coupling\footnote{The first feature is actually fed to a sigmoid function allowing to better the clustering by reducing its dependence to extreme values.}. The idea is the following: when the distance between clusters' centers and the difference in the inertia metric are high, this reveals that two clusters exist in the $4$-dimensional space, while when they are small, only one cluster should be considered. For parameters belonging to this category, the probability to reach spontaneous coupling is fixed to $1$.
    \FloatBarrier
\section*{Acknowledgements}
This work has benefited from the support of the Agence Nationale de la
Recherche through the program Investissements d’Avenir ANR-17-EURE-0001. The project leading to this publication has received funding from the French government under the “France 2030” investment plan managed by the French National Research Agency (reference: ANR-17-EURE-0020) and from Excellence Initiative of Aix-Marseille University - A*MIDEX.
\newpage

\bibliographystyle{acm}
\bibliography{biblio.bib}

@article{Calvano2020,
  doi = {10.1257/aer.20190623},
  url = {https://doi.org/10.1257/aer.20190623},
  year = {2020},
  month = oct,
  publisher = {American Economic Association},
  volume = {110},
  number = {10},
  pages = {3267--3297},
  author = {Emilio Calvano and Giacomo Calzolari and Vincenzo Denicol{\`{o}} and Sergio Pastorello},
  title = {Artificial Intelligence,  Algorithmic Pricing,  and Collusion},
  journal = {American Economic Review}
}

@article{Calvano2021,
  doi = {10.1016/j.ijindorg.2021.102712},
  url = {https://doi.org/10.1016/j.ijindorg.2021.102712},
  year = {2021},
  month = dec,
  publisher = {Elsevier {BV}},
  volume = {79},
  pages = {102712},
  author = {Emilio Calvano and Giacomo Calzolari and Vincenzo Denicol{\'{o}} and Sergio Pastorello},
  title = {Algorithmic collusion with imperfect monitoring},
  journal = {International Journal of Industrial Organization}
}

@article{Asker2022,
  doi = {10.1257/pandp.20221059},
  url = {https://doi.org/10.1257/pandp.20221059},
  year = {2022},
  month = may,
  publisher = {American Economic Association},
  volume = {112},
  pages = {452--456},
  author = {John Asker and Chaim Fershtman and Ariel Pakes},
  title = {Artificial Intelligence,  Algorithm Design,  and Pricing},
  journal = {{AEA} Papers and Proceedings}
}

@article{SanchezCartas2022,
  doi = {10.1109/access.2022.3144390},
  url = {https://doi.org/10.1109/access.2022.3144390},
  year = {2022},
  publisher = {Institute of Electrical and Electronics Engineers ({IEEE})},
  volume = {10},
  pages = {10575--10584},
  author = {J. Manuel Sanchez-Cartas and Evangelos Katsamakas},
  title = {Artificial Intelligence,  Algorithmic Competition and Market Structures},
  journal = {{IEEE} Access}
}

@article{Klein2021,
  doi = {10.1111/1756-2171.12383},
  url = {https://doi.org/10.1111/1756-2171.12383},
  year = {2021},
  month = aug,
  publisher = {Wiley},
  volume = {52},
  number = {3},
  pages = {538--558},
  author = {Timo Klein},
  title = {Autonomous algorithmic collusion: Q-learning under sequential pricing},
  journal = {The {RAND} Journal of Economics}
}

@article{johnson2023platform,
  title={Platform design when sellers use pricing algorithms},
  author={Johnson, Justin Pappas and Rhodes, Andrew and Wildenbeest, Matthijs},
  journal={Econometrica},
  volume={91},
  number={5},
  pages={1841--1879},
  year={2023},
  publisher={Econometric Society, the University of Chicago}
}

@misc{banchio_mantegazza,
  doi = {10.48550/ARXIV.2202.05946},
  url = {https://arxiv.org/abs/2202.05946},
  author = {Banchio,  Martino and Mantegazza,  Giacomo},
  title = {Adaptive Algorithms and Collusion via Coupling},
  publisher = {arXiv},
  year = {2022},
  copyright = {arXiv.org perpetual,  non-exclusive license}
}

@inproceedings{Musolff2022,
  doi = {10.1145/3490486.3538239},
  url = {https://doi.org/10.1145/3490486.3538239},
  year = {2022},
  month = jul,
  publisher = {{ACM}},
  author = {Leon Musolff},
  title = {Algorithmic Pricing Facilitates Tacit Collusion},
  booktitle = {Proceedings of the 23rd {ACM} Conference on Economics and Computation}
}

@article{Assad2020,
  doi = {10.2139/ssrn.3682021},
  url = {https://doi.org/10.2139/ssrn.3682021},
  year = {2020},
  publisher = {Elsevier {BV}},
  author = {Stephanie Assad and Robert Clark and Daniel Ershov and Lei Xu},
  title = {Algorithmic Pricing and Competition: Empirical Evidence from the German Retail Gasoline Market},
  journal = {{SSRN} Electronic Journal}
}

@article{Calvano2019,
  doi = {10.1007/s11151-019-09689-3},
  url = {https://doi.org/10.1007/s11151-019-09689-3},
  year = {2019},
  month = feb,
  publisher = {Springer Science and Business Media {LLC}},
  volume = {55},
  number = {1},
  pages = {155--171},
  author = {Emilio Calvano and Giacomo Calzolari and Vincenzo Denicol{\`{o}} and Sergio Pastorello},
  title = {Algorithmic Pricing What Implications for Competition Policy?},
  journal = {Review of Industrial Organization}
}

@misc{compte,
  doi = {10.48550/ARXIV.2304.12647},
  url = {https://arxiv.org/abs/2304.12647},
  author = {Compte,  Olivier},
  title = {Q-based Equilibria},
  publisher = {arXiv},
  year = {2023},
  copyright = {arXiv.org perpetual,  non-exclusive license}
}

@article{hettich2021algorithmic,
  title={Algorithmic collusion: Insights from deep learning},
  author={Hettich, Matthias},
  journal={Available at SSRN 3785966},
  year={2021}
}

@book{sutton2018reinforcement,
  title={Reinforcement learning: An introduction},
  author={Sutton, Richard S and Barto, Andrew G},
  year={2018},
  publisher={MIT press}
}

@article{nowe2012game,
  title={Game theory and multi-agent reinforcement learning},
  author={Now{\'e}, Ann and Vrancx, Peter and De Hauwere, Yann-Micha{\"e}l},
  journal={Reinforcement Learning: State-of-the-Art},
  pages={441--470},
  year={2012},
  publisher={Springer}
}

@article{lerer2017maintaining,
  title={Maintaining cooperation in complex social dilemmas using deep reinforcement learning},
  author={Lerer, Adam and Peysakhovich, Alexander},
  journal={arXiv preprint arXiv:1707.01068},
  year={2017}
}

@article{Tampuu2017,
  doi = {10.1371/journal.pone.0172395},
  url = {https://doi.org/10.1371/journal.pone.0172395},
  year = {2017},
  month = apr,
  publisher = {Public Library of Science ({PLoS})},
  volume = {12},
  number = {4},
  pages = {e0172395},
  author = {Ardi Tampuu and Tambet Matiisen and Dorian Kodelja and Ilya Kuzovkin and Kristjan Korjus and Juhan Aru and Jaan Aru and Raul Vicente},
  editor = {Cheng-Yi Xia},
  title = {Multiagent cooperation and competition with deep reinforcement learning},
  journal = {{PLOS} {ONE}}
}

@article{tesauro2003extending,
  title={Extending Q-learning to general adaptive multi-agent systems},
  author={Tesauro, Gerald},
  journal={Advances in neural information processing systems},
  volume={16},
  year={2003}
}

@article{hu2003nash,
  title={Nash Q-learning for general-sum stochastic games},
  author={Hu, Junling and Wellman, Michael P},
  journal={Journal of machine learning research},
  volume={4},
  number={Nov},
  pages={1039--1069},
  year={2003}
}

@article{maskin1988theory,
  title={A theory of dynamic oligopoly, II: Price competition, kinked demand curves, and Edgeworth cycles},
  author={Maskin, Eric and Tirole, Jean},
  journal={Econometrica: Journal of the Econometric Society},
  pages={571--599},
  year={1988},
  publisher={JSTOR}
}

@article{mnih2015human,
  title={Human-level control through deep reinforcement learning},
  author={Mnih, Volodymyr and Kavukcuoglu, Koray and Silver, David and Rusu, Andrei A and Veness, Joel and Bellemare, Marc G and Graves, Alex and Riedmiller, Martin and Fidjeland, Andreas K and Ostrovski, Georg and others},
  journal={nature},
  volume={518},
  number={7540},
  pages={529--533},
  year={2015},
  publisher={Nature Publishing Group}
}

@article{abada2023artificial,
  title={Artificial intelligence: Can seemingly collusive outcomes be avoided?},
  author={Abada, Ibrahim and Lambin, Xavier},
  journal={Management Science},
  year={2023},
  publisher={INFORMS}
}

@inproceedings{kennedy1995particle,
  title={Particle swarm optimization},
  author={Kennedy, James and Eberhart, Russell},
  booktitle={Proceedings of ICNN'95-international conference on neural networks},
  volume={4},
  pages={1942--1948},
  year={1995},
  organization={IEEE}
}

@article{kianercy2012dynamics,
  title={Dynamics of Boltzmann Q learning in two-player two-action games},
  author={Kianercy, Ardeshir and Galstyan, Aram},
  journal={Physical Review E},
  volume={85},
  number={4},
  pages={041145},
  year={2012},
  publisher={APS}
}

@article{banerjee2007reaching,
  title={Reaching pareto-optimality in prisoner’s dilemma using conditional joint action learning},
  author={Banerjee, Dipyaman and Sen, Sandip},
  journal={Autonomous Agents and Multi-Agent Systems},
  volume={15},
  pages={91--108},
  year={2007},
  publisher={Springer}
}

@article{forgy1965cluster,
  title={Cluster analysis of multivariate data: efficiency versus interpretability of classifications},
  author={Forgy, Edward W},
  journal={biometrics},
  volume={21},
  pages={768--769},
  year={1965}
}

@article{lloyd1982least,
  title={Least squares quantization in PCM},
  author={Lloyd, Stuart},
  journal={IEEE transactions on information theory},
  volume={28},
  number={2},
  pages={129--137},
  year={1982},
  publisher={IEEE}
}

@article{colliard2022algorithmic,
  title={Algorithmic Pricing and Liquidity in Securities Markets},
  author={Colliard, Jean-Edouard and Foucault, Thierry and Lovo, Stefano},
  journal={HEC Paris Research Paper},
  year={2022}
}

@article{singh2000convergence,
  title={Convergence results for single-step on-policy reinforcement-learning algorithms},
  author={Singh, Satinder and Jaakkola, Tommi and Littman, Michael L and Szepesv{\'a}ri, Csaba},
  journal={Machine learning},
  volume={38},
  pages={287--308},
  year={2000},
  publisher={Springer}
}

@article{xu2024mechanism,
  title={On Mechanism Underlying Algorithmic Collusion},
  author={Xu, Zhang and Zhao, Wei},
  journal={arXiv preprint arXiv:2409.01147},
  year={2024}
}

@article{dolgopolov2024reinforcement,
  title={Reinforcement learning in a prisoner's dilemma},
  author={Dolgopolov, Arthur},
  journal={Games and Economic Behavior},
  volume={144},
  pages={84--103},
  year={2024},
  publisher={Elsevier}
}

@article{axelrod1980effective,
  title={Effective choice in the prisoner's dilemma},
  author={Axelrod, Robert},
  journal={Journal of conflict resolution},
  volume={24},
  number={1},
  pages={3--25},
  year={1980},
  publisher={Sage Publications Sage CA: Los Angeles, CA}
}

@article{rubinstein1986finite,
  title={Finite automata play the repeated prisoner's dilemma},
  author={Rubinstein, Ariel},
  journal={Journal of economic theory},
  volume={39},
  number={1},
  pages={83--96},
  year={1986},
  publisher={Elsevier}
}

@article{friedman1971non,
  title={A non-cooperative equilibrium for supergames},
  author={Friedman, James W},
  journal={The Review of Economic Studies},
  volume={38},
  number={1},
  pages={1--12},
  year={1971},
  publisher={Wiley-Blackwell}
}

@article{fudenberg1986folk,
  title={The Folk Theorem in Repeated Games with Discounting or with Incomplete Information},
  author={Fudenberg, Drew and Maskin, Eric},
  journal={Econometrica},
  volume={54},
  number={3},
  pages={533--554},
  year={1986},
  publisher={Citeseer}
}

@article{askenazi2024reinforcement,
  title={Reinforcement Learning, Collusion, and the Folk Theorem},
  author={Askenazi-Golan, Galit and Cecchelli, Domenico Mergoni and Plumb, Edward},
  journal={arXiv preprint arXiv:2411.12725},
  year={2024}
}

@article{cartea2022algorithms,
  title={Algorithms can learn to collude: A folk theorem from learning with bounded rationality},
  author={Cartea, Alvaro and Chang, Patrick and Penalva, Jos{\'e} and Waldon, Harrison},
  journal={Available at SSRN 4293831},
  year={2022}
}

@article{axelrod1981emergence,
  title={The emergence of cooperation among egoists},
  author={Axelrod, Robert},
  journal={American political science review},
  volume={75},
  number={2},
  pages={306--318},
  year={1981},
  publisher={Cambridge University Press}
}

@article{miller1996coevolution,
  title={The coevolution of automata in the repeated prisoner's dilemma},
  author={Miller, John H},
  journal={Journal of Economic Behavior \& Organization},
  volume={29},
  number={1},
  pages={87--112},
  year={1996},
  publisher={Elsevier}
}

@article{cho1995perceptrons,
  title={Perceptrons Play the Repeated Prisoner's Dilemma},
  author={Cho, In-Koo},
  journal={Journal of Economic Theory},
  volume={67},
  number={1},
  pages={266--284},
  year={1995},
  publisher={Elsevier}
}

@article{busoniu2008comprehensive,
  title={A comprehensive survey of multiagent reinforcement learning},
  author={Busoniu, Lucian and Babuska, Robert and De Schutter, Bart},
  journal={IEEE Transactions on Systems, Man, and Cybernetics, Part C (Applications and Reviews)},
  volume={38},
  number={2},
  pages={156--172},
  year={2008},
  publisher={IEEE}
}

@inproceedings{banchio2022artificial,
  title={Artificial intelligence and auction design},
  author={Banchio, Martino and Skrzypacz, Andrzej},
  booktitle={Proceedings of the 23rd ACM Conference on Economics and Computation},
  pages={30--31},
  year={2022}
}

@article{brown2023competition,
  title={Competition in pricing algorithms},
  author={Brown, Zach Y and MacKay, Alexander},
  journal={American Economic Journal: Microeconomics},
  volume={15},
  number={2},
  pages={109--156},
  year={2023},
  publisher={American Economic Association 2014 Broadway, Suite 305, Nashville, TN 37203-2425}
}

@article{lambin2024less,
  title={Less than meets the eye: simultaneous experiments as a source of algorithmic seeming collusion},
  author={Lambin, Xavier},
  journal={Available at SSRN 4498926},
  year={2024}
}

@article{compte2002capacity,
  title={Capacity constraints, mergers and collusion},
  author={Compte, Olivier and Jenny, Frederic and Rey, Patrick},
  journal={European Economic Review},
  volume={46},
  number={1},
  pages={1--29},
  year={2002},
  publisher={Elsevier}
}

@book{tirole1988theory,
  title={The theory of industrial organization},
  author={Tirole, Jean},
  year={1988},
  publisher={MIT press}
}

@article{ivaldi2003economics,
  title={The economics of tacit collusion},
  author={Ivaldi, Marc and Jullien, Bruno and Rey, Patrick and Seabright, Paul and Tirole, Jean},
  year={2003},
  publisher={IDEI Working Paper}
}
\end{document}